\documentclass[useAMS,usenatbib]{mn2e}
\usepackage{graphicx}

\title[High-redshift AGN in the CDF-S]{The high-redshift ($z>3$) AGN population in the 4 Ms \textit{Chandra} Deep Field South}
\author[F. Vito et al.]
{F. Vito$^{1,2}$\thanks{E-mail: fabio.vito@unibo.it},
C. Vignali$^{1,2}$,
R. Gilli$^{2}$,
A. Comastri$^{2}$, 
K. Iwasawa$^{3}$,
W.N. Brandt$^{4,5}$,
\newauthor
D.M. Alexander$^{6}$,
M. Brusa$^{7}$,
B. Lehmer$^{8,9}$,
F.E. Bauer$^{10,11}$, 
D.P. Schneider$^{4,5}$,
\newauthor
Y.Q. Xue$^{12}$,
B. Luo$^{4,5}$
\\ \\
$^{1}$ Dipartimento di Astronomia, Universit\`a degli Studi di Bologna, Via Ranzani 1, 40127 Bologna, Italy \\
$^{2}$ INAF -- Osservatorio Astronomico di Bologna, Via Ranzani 1, 40127 Bologna, Italy\\
$^{3}$ ICREA and Institut de Ciències del Cosmos (ICC), Universitat de Barcelona (IEEC-UB), Mart\'{i} i Franqu\`{e}s 1, 08028, Barcelona\\
$^{4}$ Department of Astronomy and Astrophysics, The Pennsylvania State University, University Park, PA 16802, USA\\
$^{5}$ Institute for Gravitation and the Cosmos, The Pennsylvania State University, University Park, PA 16802, USA\\
$^{6}$ Department of Physics, University of Durham, South Road, Durham, DHI 3LE, UK\\
$^{7}$ Max Planck Institut f\"{u}r Extraterrestische Physik, Postfach, 1312, 85741, Garching bei M\"{u}nchen, Germany\\
$^{8}$ The Johns Hopkins University, Homewood Campus, Baltimore, MD 21218, USA\\
$^{9}$ NASA Goddard Space Flight Centre, Code 662, Greenbelt, MD 20771, USA\\
$^{10}$ Pontificia Universidad Cat\'{o}lica de Chile, Departamento de Astronom\'{\i}a y Astrof\'{\i}sica, Casilla 306, Santiago 22, Chile\\
$^{11}$ Space Science Institute, 4750 Walnut Street, Suite 205, Boulder, Colorado 80301, USA\\
$^{12}$ Key Laboratory for Research in Galaxies and Cosmology, Department of Astronomy, University of Science and Technology of\\ China, Chinese Academy of Sciences, Hefei, Anhui, 230026, China\\
}

\begin{document}

\date{Accepted 2012 September 18.  Received 2012 September 15; in original form 2012 August 3}

\graphicspath{{spectra/}}

\pagerange{\pageref{firstpage}--\pageref{lastpage}} \pubyear{2012}

\maketitle

\label{firstpage}

\begin{abstract}
We present results from a spectral analysis of a sample of high-redshift ($z>3$) X-ray selected AGN in the 4 Ms \textit{Chandra} Deep Field South (CDF-S), the deepest X-ray survey to date. The sample is selected using the most recent spectroscopic and photometric information available in this field. It consists of 34 sources with median redshift $z=3.7$, 80 median net counts in the 0.5-7 keV band and median rest-frame absorption-corrected luminosity $L_{2-10\,\rmn{keV}}\approx1.5\times10^{44}\rmn{erg}\,\rmn{s^{-1}}$. Spectral analysis for the full sample is presented and the intrinsic column density distribution, corrected for observational biases using spectral simulations, is compared with the expectations of X-ray background (XRB) synthesis models. We find that $\approx57$ per cent of the sources are highly obscured ($N_H>10^{23}\rmn{cm^{-2}}$). Source number counts in the $0.5-2\,\rmn{keV}$ band down to flux $F_{0.5-2\,\rmn{keV}}\approx4\times10^{-17}\rmn{erg}\,\rmn{s^{-1}cm^{-2}}$ are also presented. 
Our results are consistent with a decline of the AGN space density at $z>3$ and suggest that, at those redshifts, the AGN obscured fraction is in agreement with the expectations of XRB synthesis models. 
\end{abstract}


\section{Introduction}\label{sec1}
The discovery of Super Massive Black Holes (SMBH) at $z>6$ with masses of the order of $10^{9}\,\rmn{M_\odot}$ (e.g. \citealt{Mortlock11}, the redshift record $z=7.085$; \citealt{Willott03, Fan06, Fan06b, Willott09}) implies that SMBH were already in place when galaxies were only a few hundred Myr old. The formation of these massive objects and the physical conditions allowing a rapid and early growth are challenging problems. The origin of the SMBH seeds is also uncertain. Two classes of formation models are currently under discussion: light seeds, perhaps the remnants of Pop III stars ($\sim10^{2}\,\rmn{M_\odot}$; \citealt*{Madau01, Alvarez09, Volonteri10}, see also \citealt{Whalen11}), possibly evolved as a ``quasi-star"\citep[]{Begelman10}, and heavy seeds, formed by monolithic collapse of massive gas clouds \citep[$10^{4-5}\,\rmn{M_\odot}$; e.g.][]{Ball11}. Whatever the nature of the seed is, continuous Eddington-limited or even super-Eddington accretion onto these objects is required to reach the 
observed masses.

Such a high accretion rate requires a large amount of accreting material which, at the same time, may act as a screen against the emitted radiation \citep[e.g.][]{Hopkins08}. Supporting observational evidence was found for an increasing fraction of obscured Active Galactic Nuclei (AGN) from $z=0$ to $z\approx2-3$ (\citealt{LaFranca05,Ballantyne06, Treister06, Hasinger08}; but see also \citealt{Gilli10}, who dispute these results). At these redshifts, a large amount of gas is available in the host galaxy. The nuclear activity could be triggered by the gas falling towards the centre, when a ``wet" (i.e. gas-rich) merging event occurs. After an initial obscured ($N_H>10^{22}\rmn{cm^{-2}}$) accretion phase, which in some models can be Compton-thick ($N_H>10^{24}\rmn{cm^{-2}}$), the emitted radiation sweeps out the gas, revealing itself and acting as a negative feedback towards the star formation in the host galaxy \citep[e.g.][]{Brusa05, Alexander10, Feruglio10, Sturm11, Maiolino12}. The evolution of obscuration,
 in 
particular 
at high redshift, when most of the mass-accretion was taking place, is a fundamental observable to probe this framework. 

 The AGN comoving space density in the optical and X-ray band is not constant over time: there are many more AGN at $z>1$ than in the local Universe \citep[e.g.][]{Hasinger05}. The evolution of the AGN population can be described by a phenomenological model known as Luminosity Dependent Density Evolution model \citep*[LDDE model; e.g. ][]{Hasinger05, Brandt10}: according to this model, the comoving space density of more luminous AGN ($L_{X}>10^{45} \,\rmn{erg\, s^{-1}}$) peaks at $z\sim2-3$ and then decreases exponentially to $z\sim6$; for less luminous ones ($L_{X}\la10^{45} \,\rmn{erg\, s^{-1}}$) the peak arises at more recent times ($z\sim1-1.5$) and then their density slowly decreases till the highest redshifts so far probed ($z\sim3$). Therefore, more massive black holes formed and grew earlier than smaller ones. This sort of ``cosmic downsizing" is similar to that found for galaxies \citep[e.g.][]{Cowie99, Thomas05, Damen09}.

This scenario is, however, far from being complete. Particularly, the behaviour at higher redshifts ($z\geq3$) is not constrained (especially at $L_X<10^{44}\rmn{erg}\,\rmn{s^{-1}}$), implying severe uncertainties in AGN evolution during the early ages of the Universe. Furthermore, the space density and the role of heavily obscured AGN are largely unknown beyond the local Universe. In order to place constraints on the obscured AGN fraction and its evolution history, observables, such as the spectral properties and the number counts, are required to allow comparisons with evolutionary theoretical models.

Optical surveys are not suitable for these tasks since they are severely biased against obscuration. Deep X-ray surveys are required, because they are less affected by obscuration and can reach fluxes faint enough to sample the bulk of the high-redshift AGN population. \cite{Trichas12} presented the largest (78 objects) sample of $z > 3$ X-ray selected AGN, using data from the Chandra Multi-wavelength Project (ChaMP) survey. A few studies have been performed on high-redshift AGN exploiting the deep surveys in the Cosmological Evolution Survey (COSMOS) field carried out with \textit{XMM-Newton} \citep[]{Brusa09} and \textit{Chandra} \citep[]{Civano11}, and they were limited to relatively bright X-ray fluxes ($F_{0.5-2\mathrm{keV}}\ga10^{-15}$ and $3\times10^{-16}\,\rmn{erg\,cm^{-2}s^{-1}}$, respectively) and high $2-10\, \rmn{keV}$ luminosities ($L_{2-10\, \rmn{keV}}>10^{44.2}\,\rmn{erg}\,\rmn{ s^{-1}}$ and $10^{43.5}\,\rmn{erg}\,\rmn{ s^{-1}}$, respectively). Only recently, thanks to the 4 Ms \textit{Chandra} Deep Field South (CDF-S, \citealt{Xue11}), the deepest X-ray survey to date, very faint fluxes (down to $F\approx3.2\
\times10^{-17}/9.1\times10^{-18}/5.5\times10^{-17}\,\rmn{erg\,cm^{-2}s^{-1}}$ in the $0.5-8/0.5-2/2-8\,\rmn{keV}$ band) could be reached. Early results on $z>3$ population from the 4 Ms CDF-S data have been presented in \cite{Fiore12}. \cite{Lehmer12} also presented the source counts of the $z>3$ population as part of a more comprehensive work on logN-logS of the X-ray population.

In this work, we analyse data from the 4 Ms CDF-S. In particular, we have selected a sample of ``bona fide" high-redshift ($z>3$) AGN using the most recent spectroscopic and photometric information available, and performed an X-ray spectral study to understand their properties and, possibly, constrain the early phases of AGN evolution. The spectral properties, column-density distribution and number counts of the sample are derived, discussed and compared with previous works and model predictions.

Throughout this work we assume a $\Omega_{m}=0.3$, $\Omega_{\Lambda}=0.7$ and $H_0 = 70 \,\rmn{km\,s^{-1}Mpc^{-1}}$ Universe \citep{Spergel03}. The errors on the spectral parameters correspond to the 90 per cent confidence level for one parameter of interest \citep{Avni76}.

\section[]{The $\lowercase{z}>3$ AGN sample}\label{sec2}
We assembled a sample of X-ray selected AGN at $z>3$ in the 4 Ms CDF-S  on the basis of both spectroscopic and photometric redshifts.  Spectroscopic redshifts for $z>3$ AGN were collected from \citet{Szokoly04}, \citet{Vanzella08}, \citet{Popesso09}, \citet{Silverman10}, \citet{Vanzella10}, \citet{Wuyts09} and E. Vanzella (private communication\footnote{These spectroscopic redshifts are based on a re-analysis of VIsible MultiObject Spectrograph (VIMOS) and FOcal Reducer and low dispersion Spectrograph (FORS2) spectra.}). A quality flag (``secure/insecure") is associated to them in \cite{Xue11}, except for spectroscopic redshifts from \cite{Vanzella10}, \citet{Wuyts09} and E. Vanzella (private communication), which are not used by \cite{Xue11} and for which we retrieved the original quality information. Photometric redshifts were gathered from \citet[which also provides secondary solutions]{Luo10}, \citet{Cardamone10}, \citet{Rafferty11}, \citet{Santini09}, \citet{Wuyts08}, \citet{Taylor09} and \citet{Dahlen10}. 

\begin{figure*}
\begin{center}
\includegraphics[width=160mm,keepaspectratio]{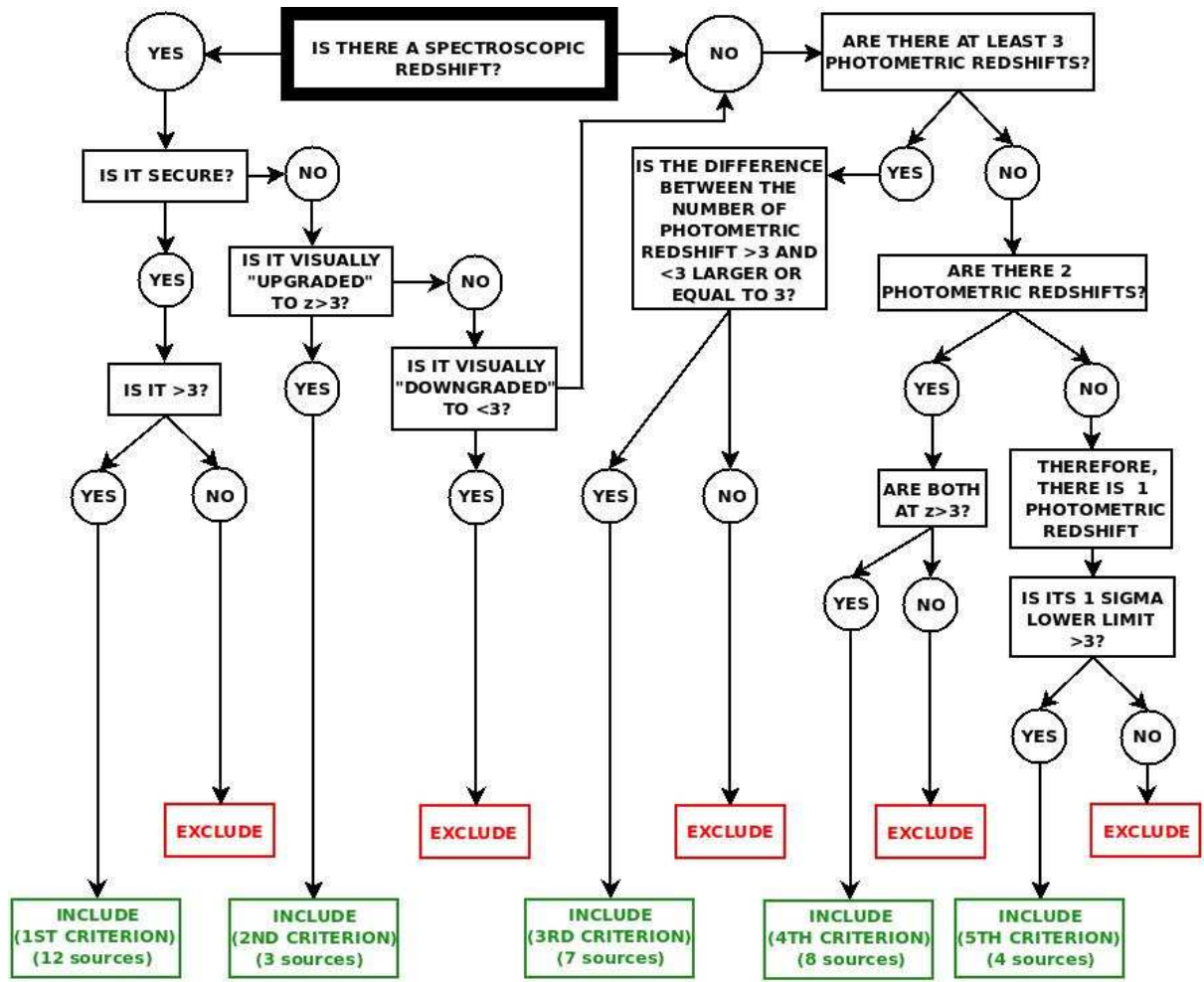} 
\end{center}
\caption{Selection criteria applied to the 96 pre-selected sources (which have at least one spectroscopic or photometric redshift $z>3$) to be included in the high-redshift sample, visualised as a flow chart. The starting block has bold sides.}
\label{fig1}
 \end{figure*}

We first cross-matched the X-ray positions of all sources in the 4 Ms CDF-S main catalogue \citep[740 sources;][]{Xue11} with the positions of the sources in the catalogues which are not collected by \cite{Xue11} using a 1 arcsec matching radius. Considering the other catalogues, we assumed the counterparts reported by \cite{Xue11}, which used a likelihood-ratio matching technique. We pre-selected a sample of high-redshift AGN candidates, including all sources which have a spectroscopic or photometric redshift $z>3$ in at least one of the considered catalogues (96 sources). Given the large amount of available information, we built up a set of criteria to select the final high-redshift sample. A source is included in the sample if it fulfils one of the following requirements:
\begin{enumerate}
\renewcommand{\theenumi}{(\arabic{enumi})}
 \item A spectroscopic redshift flagged as ``secure" $z_{spec,s}>3$ is provided.
 \item A spectroscopic redshift flagged as ``insecure" $z_{spec,i}>3$ is available and it is upgraded to ``secure" by re-analysing the optical spectrum, checking the quality and reliability of the identified spectral features. 
 \item $N_{phot,>3} - N_{phot,<3} \geq 3$, where $N_{phot,>3}$ is the number of photometric redshifts larger than 3 and $N_{phot,<3}$ is the number of photometric redshifts lower than 3, when no secure spectroscopic redshift and more than two photometric redshifts are available.
\item When no secure spectroscopic redshift and two photometric redshifts are available, both photometric redshifts are larger than 3.
\item When no secure spectroscopic redshift and only one photometric redshift is available, its $1\sigma$ lower bound (provided by the original catalogue) is larger than 3.
\end{enumerate}
The selection procedure is shown in Fig.~\ref{fig1} as a flow chart. We will discuss in \S~\ref{discuss} the effects of relaxing the selection criteria on the sample size.

The final sample consists of 34 sources (see Tab.~\ref{tab1}). A spectroscopic redshift is adopted for 15 sources (12 and 3 are included in the high-redshift sample on the basis of the first and second criterion, respectively) and a photometric redshift is adopted for 19 sources (7/8/4 sources are included in the sample on the basis of the third/fourth/fifth criterion, respectively). XID 392 has no redshift information in \citet{Xue11} and a photometric redshift $z=6.22$ from \citet{Santini09}, but no error on the redshift is provided. Therefore, we conservatively excluded this source from the sample. The sample presented in this work is a factor $\sim2$ smaller than that discussed in \cite{Lehmer12}. This difference will be discussed in more details in \S~\ref{discuss}.

When more than one redshift per source is available, we almost entirely followed the choice of \citet{Xue11}. Briefly, the adopted redshift in this work is, in order of preference:  secure (or upgraded to secure) spectroscopic redshift, photometric redshift from \citet{Luo10}, photometric redshift from \citet{Cardamone10}. XID 331 has no redshift information in the catalogues collected by \citet{Xue11}, but the inclusion of the other above-mentioned catalogues  allows that source to be included in the sample. We assumed photometric redshift from \citet{Dahlen10} for it. 
No spectroscopic redshift is reported in \citet{Xue11} for XID 235, 458, 528 and 386; redshift for these sources were obtained by
E. Vanzella (XID 235 and 458, private communication), \citet[XID 528]{Vanzella10} and \citet[XID 386]{Wuyts09}. A high quality flag is associated to all these redshift in the original works; therefore, we consider them as ``secure".

\begin{figure}
\includegraphics[width=80mm,keepaspectratio]{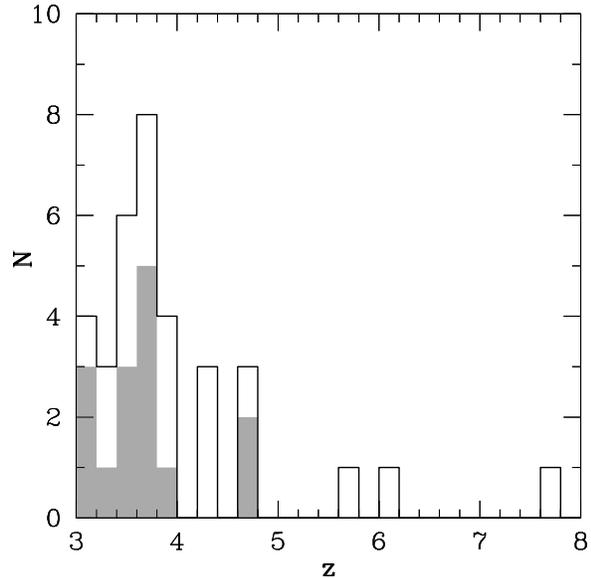} 
\caption{Redshift distribution of all 34 sources (solid line) and sources with spectroscopic redshift (shaded histogram) in the sample. }
\label{fig2}
\end{figure}

The redshift distribution of the sample is shown in Fig.~\ref{fig2}. The mean and median redshifts are 4.0 and 3.7, respectively. Most of the sources with a spectroscopic redshift (13 out of 15) are at $z<4$. We note the presence of 3 sources at $z>5$, for which we investigated in more detail the redshift information. Two of them (XID 139 and 197) have only one photometric redshift \citep[$z=5.729$ and 6.071, respectively, from][which also provides a secondary solution of $z=4.385$ for XID 197]{Luo10}, while XID 485 has two different photometric redshifts ($z=7.62$ from \citealt{Luo10}, with secondary solution $z=3.309$, and $z=4.42$ from \citealt{Santini09}). These objects have no optical counterparts in the Great Observatories Origins Deep Survey (GOODS; XID 139 and 485) or Galaxy Evolution from Morphologies and SEDs (GEMS; XID 197, which is not in the GOODS field) observations, while they are significantly detected by the Infrared Array Camera (IRAC) in the near-IR. Thus, a high-redshift ($z>3$) solution appears probable.

\section[]{Data analysis}\label{sec3}
\subsection{Spectral extraction procedure}
\textit{Chandra} data products for each of the 54 observations of the CDF-S are publicly available in the \textit{Chandra} Data Archive\footnote{http://cxc.harvard.edu/cda/}. We retrieved all data products (in particular, the evt2 files; i.e., the event list filtered on the good time intervals and status bits), but those related to the first two observations \citep[for which the focal-plane temperature was $-110^{\circ}$; ][]{Luo08,Xue11}, and used them for spectral analysis. The sources were located in each of the 52 CDF-S images according to the coordinates provided by the main-source catalogue \cite[]{Xue11}.

We selected the extraction regions for the source and background using SAOImage DS9\footnote{http://hea-www.harvard.edu/RD/ds9/}\citep{Joye03}. Despite the \textit{Chandra} PSF distortion with the off-axis angle, we always chose circular regions, provided that most of the source counts are included in the selected regions (probably this choice slightly lowers the signal-to-noise ratio). Small radii are required for faint sources in order to include a low number of background counts and therefore to increase the signal-to-noise ratio. Larger extraction radii can be used for bright sources. Also the position of the source in the field of view (f.o.v.) must be taken into account, because of the degradation of the PSF with the off-axis angle. For each source, we defined the same radius for the extraction region in all observations, considering the mean off-axis angle, provided by the main catalogue \citep[]{Xue11}, and the relation between it and the encircled energy radius.

The off-axis angles for the sources in the sample are in the range of $1.42-10.1\,\rmn{arcmin}$. We chose the extraction radius for each source to obtain an acceptable trade-off between the number of counts and the signal-to-noise ratio. As a general rule, for sources at off-axis angles $\la4$ arcmin we defined the extraction radius corresponding to an encircled energy fraction (EEF) of 90 per cent at 1.49 keV. For sources at a large off-axis angle ($\ga7$ arcmin), we considered radii in a range corresponding to 50-90 per cent EEF at 1.49 keV. For sources at intermediate off-axis angles (4-7 arcmin) we tried different extraction radii and chose those which provided the best trade-off between source and background counts. In a few cases we were forced to use smaller radii to avoid the contamination by sources close to the one we were considering. As a result, the extraction radii for the sample are in the range of $1.5-5\,\rmn{arcsec}$. Regarding the background, we chose extraction regions as close as 
possible to the source, preferably on the same chip, in areas where no other source was detected. In particular, we excluded all the regions around detected sources with radii a factor $\ge1.2$ larger than the radii corresponding to 90\% EEF at 6.4 keV (which are a factor $>1.5$ larger than the radii corresponding to 90\% EEF at 1.49 keV), in order to avoid contamination.

We used CIAO v4.3\footnote{http://cxc.harvard.edu/ciao/} to extract source and background spectra and to create the response files for each observation. In particular, we used the \textit{specextract} tool which contains some improvements with respect to similar scripts in previous CIAO versions (e.g., it applies the energy-dependent aperture correction to the ARF file, accounting for the fraction of the PSF enclosed by a region). The spectra and response files of each source obtained for each observation were then added using tools provided in the FTOOLS package\footnote{http://heasarc.nasa.gov/ftools/ftools\_menu.html}(i.e. \textit{mathpha}, \textit{addarf} and \textit{addrmf}; \citealt{Blackburn95}), weighted by the exposure time of the individual observations.

\begin{figure}
\includegraphics[width=80mm,keepaspectratio]{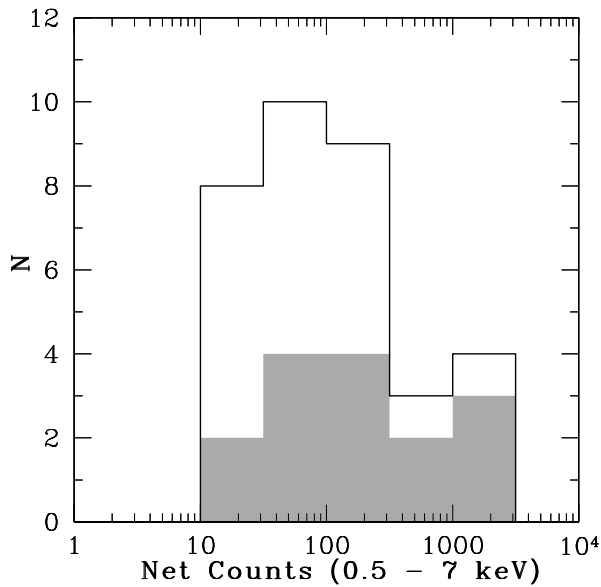}
\caption{Full-band ($0.5-7\,\rmn{keV}$) net counts distribution of all 34 
sources (solid line) and sources with spectroscopic redshift (shaded histogram) in the sample.}
\label{fig3}
\end{figure}

The distribution of full-band ($0.5-7\,\rmn{keV}$) net (i.e. background subtracted) counts (Fig.~\ref{fig3}) is peaked near 100 counts (the mean and median values are 275 and 80, respectively). As expected, the bulk of the objects has a very low number of net counts: $\sim74$ per cent of sources in the sample have less than 200 net counts. In order to maximise the signal-to-noise ratio (SNR), in most cases we fitted the spectra in a narrower energy range ($0.5-5 \,\rmn{keV}$), given that in the $5-7\,\rmn{keV}$ band the background is high. All the spectra were fitted using the Cash statistic (\citeauthor{Cash79} \citeyear{Cash79}) to estimate the best-fitting parameters.

\subsection{Spectral analysis}\label{sec4}
The spectral fitting was performed with XSPEC v12.5.1n\footnote{http://heasarc.nasa.gov/xanadu/xspec/} \citep{Arnaud96}. We assumed an absorbed power-law, parametrized by XSPEC models \textit{powerlaw} and \textit{zwabs}, as the starting spectral model. Galactic absorption was included assuming a column density in the range $(6-9)\times10^{19}\rmn{cm^{-2}}$ \citep[]{Kalberla05}.

The mean photon index is $\langle\Gamma\rangle=1.51$ with dispersion $\sigma=0.28$ for the 8 brightest sources ($>250$ net counts) and $\langle\Gamma\rangle=1.63$ with $\sigma=0.89$ for the whole sample. The average photon index is slightly flatter than what is typically measured in AGN samples ($\Gamma=1.8$; e.g. \citealt{Tozzi06, Turner97}), but we emphasize that we are sampling the rest-frame $2-20\,\rmn{keV}$ band, and the likely presence of a reflection hump could flatten the observed X-ray spectrum. However, the measured value is statistically consistent with  $\Gamma=1.8$. Reasonable accuracy was obtained for the best-fitting parameters only for the 8 brightest sources (with a mean error on $\Gamma$ and $N_H$ of 14 and 45 per cent, respectively). The spectral parameters were not well constrained for almost all the other sources (error on $\Gamma$ of 117 per cent; only upper limits on $N_H$ for most objects) because of the poor spectral quality. 

The difficulties in simultaneously constraining $\Gamma$ and $N_H$ for sources with few counts can be explained considering that these two spectral parameters have a significant degree of degeneracy. The problem is further complicated by the redshifted absorption cut-off, which moves outside the \textit{Chandra} band-pass at $z\ga3$ even for significant degrees of obscuration, introducing further uncertainties. 

To better constrain the column density, we fixed the photon index to $\Gamma=1.8$ and repeated the spectral analysis on all the sources. This photon index is a widely used value for sources with low counting statistics, being considered the typical slope for AGN power-law emission in X-rays \citep[e.g. ][]{Turner97,Tozzi06}. The best-fitting parameters, fluxes and intrinsic luminosities derived using this spectral model are reported in Tab.~\ref{tab1}. Similar results are obtained fitting simultaneously the source and background spectra. All the X-ray spectra of the sources, fitted using this model, are shown in Appendix B. The effects of assuming a flatter photon index ($\Gamma=1.6$) will be discussed in \S~\ref{nh} and \S~\ref{discuss}.

\begin{table*} 
\caption{Main information and spectral parameters of the 34 high-redshift AGN.}
\label{tab1}
\begin{tabular}{rrrrrrrrrrrr }
\hline
  \multicolumn{1}{c}{XID} &
  \multicolumn{1}{c}{RA[J2000]} &
  \multicolumn{1}{c}{DEC[J2000]} &
  \multicolumn{1}{c}{z} &
  \multicolumn{1}{c}{r} &
  \multicolumn{1}{c}{C} &
  \multicolumn{1}{c}{$N_H$} &
  \multicolumn{1}{c}{CSTAT/DOF} &
  \multicolumn{1}{c}{$F_{0.5-2\,\rmn{keV}} $} &
  \multicolumn{1}{c}{$F_{2-10\,\rmn{keV}} $} &
  \multicolumn{1}{c}{$L_{2-10\,\rmn{keV}} $} &
  \multicolumn{1}{c}{COUNTS} \\
  
  \multicolumn{1}{c}{(1)} &
  \multicolumn{1}{c}{(2)} &
  \multicolumn{1}{c}{(3)} &
  \multicolumn{1}{c}{(4)} &
  \multicolumn{1}{c}{(5)} &
  \multicolumn{1}{c}{(6)} &
  \multicolumn{1}{c}{(7)} &
  \multicolumn{1}{c}{(8)} &
  \multicolumn{1}{c}{(9)} &
  \multicolumn{1}{c}{(10)} &
  \multicolumn{1}{c}{(11)} &
  \multicolumn{1}{c}{(12)} \\
\hline
27  & 52.96054 & -27.87706 & 4.385 & P8 & 5 &$68^{+41}_{-29}$   & 71.7/113  & 2.45E-16   & 1.76E-15    & 2.80E44 & 77   \\[1ex]
100 & 53.01658 & -27.74489 & 3.877 & P8 & 3 &$38_{-24}^{+27}$   & 69.0/92   & 1.48E-16   &$<$7.55E-16  & 8.94E43 & 67   \\[1ex]
107 & 53.01975 & -27.66267 & 3.808 & P9 & 4 &$<71$              & 51.6/62   & 4.80E-16   &$<$1.40E-15  & 1.52E44 & 36   \\[1ex]
132 & 53.03071 & -27.82836 & 3.528 & P8 & 3 &$13^{+16}_{-11}$   & 69.0/65   & 1.13E-16   &$<$3.39E-16  & 3.12E43 & 57   \\[1ex]
\smallskip
139 & 53.03496 & -27.67975 & 5.729 & P8 & 5 &$96^{+33}_{-35}$   & 92.8/101  & 5.71E-16   & 3.64E-15    & 1.02E45 & 111  \\[1ex]
150 & 53.03996 & -27.79847 & 3.337 & P9 & 3 &$45^{+49}_{-39}$   & 17.3/31   &$<$2.24E-17 & 1.58E-16    & 1.40E43 & 26   \\[1ex]
170 & 53.04746 & -27.87047 & 3.999 & P8 & 4 &$35_{-4}^{+4}$     & 308/359   & 1.63E-15   & 7.49E-15    & 9.43E44 & 1472 \\[1ex]
197 & 53.05771 & -27.93344 & 6.071 & P8 & 5 &$86^{+31}_{-30}$   & 88.2/124  & 6.24E-16   & 3.27E-15    & 1.02E45 & 134   \\[1ex]
235 & 53.07029 & -27.84564 & 3.712 & S7 & 1 &$<36$              & 33.0/27   & 4.29E-17   & 7.95E-17    & 8.06E42 & 27   \\[1ex]
\smallskip
262 & 53.07854 & -27.85992 & 3.66  & S1 & 1 &$85_{-20}^{+22}$   & 123.6/165 & 1.20E-16   & 1.53E-15    & 1.63E44 & 208  \\[1ex]
283 & 53.08467 & -27.70811 & 3.204 & P8 & 4 &$55^{+39}_{-22}$   & 58.0/80   & 1.04E-16   & 9.54E-16    & 7.93E43 & 77   \\[1ex]
285 & 53.08558 & -27.85822 & 4.253 & P8 & 3 &$<6$               & 29.7/31   & 5.71E-17   &$<$9.13E-17  & 1.24E43 & 18   \\[1ex]
331 & 53.10271 & -27.86061 & 3.78  & P10 & 4 &$<23$              & 48.1/36   & 4.61E-17   &$<$7.39E-17  & 7.71E42 & 22   \\[1ex]
371 & 53.11158 & -27.76789 & 3.101 & P8 & 4 &$35^{+11}_{-10}$   & 98.1/118  & 1.22E-16   & 7.70E-16    & 5.70E43 & 146  \\[1ex]
\smallskip
374 & 53.11204 & -27.86072 & 3.724 & P9 & 4 &$213^{+284}_{-114}$& 32.7/34   &$<$3.59E-18 & 2.95E-16    & 4.14E43 & 22   \\[1ex]
386 & 53.11796 & -27.73439 & 3.256 & S6 & 1 &$<7$               & 62.6/64   & 8.85E-17   &$<$1.42E-16  & 1.06E43 & 40   \\[1ex]
403 & 53.12196 & -27.93883 & 4.762 & S2 & 1 &$185_{-76}^{+148}$ & 66.9/73   &$<$5.46E-17 & 1.30E-15    & 2.68E44 & 36   \\[1ex]
412 & 53.12442 & -27.85169 & 3.7   & S1 & 1 &$82_{-12}^{+12}$   & 208.0/225 & 2.08E-16   & 2.46E-15    & 2.85E44 & 373  \\[1ex]
458 & 53.13854 & -27.82128 & 3.474 & S7 & 1 &$93^{+74}_{-70}$   & 14.7/25   & 1.59E-17   & 2.52E-16    & 2.64E43 & 27   \\[1ex]
\smallskip
485 & 53.14658 & -27.87103 & 7.620 & P8 & 4 &$406^{+192}_{-114}$& 74.2/94   & 5.75E-17   & 8.86E-16    & 5.00E44 & 87   \\[1ex]
521 & 53.1585  & -27.73372 & 3.417 & P8 & 4 &$<23$              & 47.6/50   & 8.91E-17   &$<$2.40E-16  & 2.05E43 & 30   \\[1ex]
528 & 53.16158 & -27.85606 & 3.951 & S5 & 1 &  $74^{+24}_{-24}$ & 81.6/111  & 9.56E-17   & 8.93E-16    & 1.16E44 & 126  \\[1ex]
546 & 53.16533 & -27.81419 & 3.064 & S1 & 1 &$52_{-4}^{+4}$     & 331.1/355 & 6.75E-16   & 6.26E-15    & 4.73E44 & 1072 \\[1ex]
556 & 53.17012 & -27.92975 & 3.528 & P8 & 3 &$97_{-10}^{+9}$    & 285.6/293 & 7.59E-16   & 1.07E-14    & 1.14E45 & 718  \\[1ex]
\smallskip
563 & 53.17442 & -27.86742 & 3.61  & S1 & 1 &$6_{-2}^{+2}$      & 257.6/307 & 2.07E-15   & 4.66E-15    & 4.45E44 & 1084 \\[1ex]
573 & 53.1785  & -27.78411 & 3.193 & S1 & 2 &$3_{-2}^{+2}$      & 209.2/218 & 8.10E-16   & 1.65E-15    & 1.19E44 & 609  \\[1ex]
588 & 53.18467 & -27.88103 & 3.471 & S1 & 1 &$<2$               & 115.5/153 & 6.45E-16   & 1.03E-15    & 8.41E43 & 281  \\[1ex]
642 & 53.20821 & -27.74994 & 3.769 & P9 & 3 &$<13$              & 55.7/65   & 9.17E-17   &$<$1.48E-16  & 1.54E43 & 18   \\[1ex]
645 & 53.20933 & -27.88119 & 3.47  & S3 & 2 &$15_{-2}^{+2}$     & 371.4/384 & 3.16E-15   & 1.03E-14    & 9.24E44 & 2053 \\[1ex]
\smallskip
651 & 53.21529 & -27.87033 & 4.658 & P8 & 3 &$151^{+39}_{-35}$  & 115.8/132 & 1.33E-16   & 2.38E-15    & 4.64E44 & 146  \\[1ex]
674 & 53.24004 & -27.76361 & 3.082 & S3 & 1 &$<7$               & 120.3/146 & 6.95E-16   & 1.39E-15    & 8.56E43 & 199  \\[1ex]
700 & 53.2625  & -27.86308 & 4.253 & P8 & 5 &$18^{+16}_{-14}$   & 73.8/90   & 7.43E-16   & 2.23E-15    & 3.11E44 & 102  \\[1ex]
717 & 53.28    & -27.79892 & 4.635 & S4 & 2 &$87_{-51}^{+65}$   & 47.2/73   & 1.21E-16   & 1.02E-15    & 1.85E44 & 44   \\[1ex]
730 & 53.29587 & -27.79317 & 3.724 & S4 & 1 &$60_{-39}^{+49}$   & 59.6/83   &$<$1.81E-16 &$<$1.46E-15  & 1.65E44 & 37   \\[1ex]
\hline                                    

\end{tabular}

(1) Identification number from \citet{Xue11}; (2) Right ascension and (3) declination from \citet{Xue11}; (4) adopted redshift (see \S~\ref{sec2}); (5) reference for the adopted redshift -  S1, S2, S3, S4, S5, S6 and S7: spectroscopic redshift from \citet{Szokoly04}, \citet{Vanzella08}, \citet{Popesso09}, \citet{Silverman10}, \citet{Vanzella10}, \citet{Wuyts09} and E. Vanzella (private communication), respectively; P8, P9 and P10: photometric redshift from \citet{Luo10}, \citet{Cardamone10} and \citet[weighted solution]{Dahlen10}, respectively; (6) criterion satisfied for source inclusion in the high-redshift sample (see \S~\ref{sec2}); (7) best-fitting $N_H$ as derived in \S~\ref{sec4}, in units of $10^{22}\,\rmn{cm^{-2}}$. We define $N_H$ to be constrained if the lower limit (at the 90 per cent c.l.) on the best-fitting value is larger than zero, otherwise we report its upper limit. (8) best-fit value of Cash statistic over degrees of freedom; (9) soft-band ($0.5 - 2 \,\rmn{keV}$) and (
10) hard-band ($2 - 10 \,\rmn{keV}$) flux or upper limits from the best-fitting spectral model, in units of $\,\rmn{erg\,cm^{-2}s^{-1}}$; (11) intrinsic (i.e. absorption-corrected) rest-frame $2-10\,\rmn{keV}$ luminosity, in units of $\,\rmn{erg\,s^{-1}}$ (typical errors range from $\sim10$ per cent for the brightest sources to a factor of $\sim2$ for the faintest ones, fixing the redshift to the value reported in column 4); (12) full-band ($0.5-7\,\rmn{keV}$) net (i.e. background-subtracted) counts, as obtained from the spectral extraction. The full-band net-count number is only representative, being the energy range considered during the spectral fitting narrower, in most cases, in order to maximise signal-to-noise ratio (SNR).
\end{table*}

\begin{figure}
\includegraphics[width=80mm,keepaspectratio]{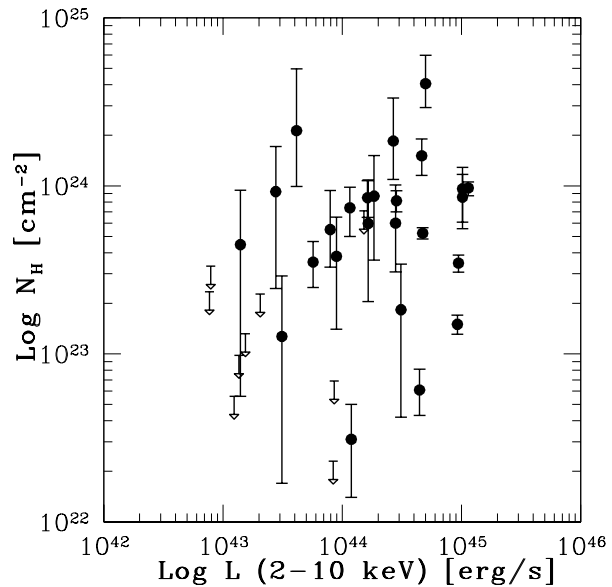}
\caption{Best-fitting $N_H$ as a function of the hard-band intrinsic luminosity for all the sources in the sample. Filled circles represent constrained values, while upper limits are reported as downward-pointing arrows.}
\label{fig4}
\end{figure}

\begin{figure}
\includegraphics[width=80mm,keepaspectratio]{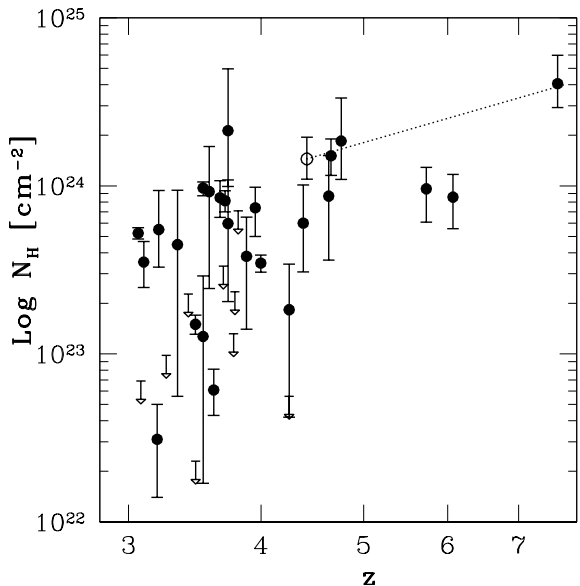}
\caption{Best-fitting $N_H$ as a function of redshift for all the sources in the sample. Filled circles represent constrained values, while upper limits are reported as downward-pointing arrows. The dashed line connects the best-fitting $N_H$ of XID 485 at $z=7.62$, as reported in Tab.~\ref{tab1}, with the best-fitting value which would be derived if it were placed at $z=4.42$ (open circle). See text for details.}
\label{fig5}
\end{figure}

An anti-correlation between the column density and the AGN intrinsic luminosity is reported in several works (e.g. \citealt{Steffen03}, \citealt{Ueda03}, \citealt{Hasinger04}, \citealt{LaFranca05}, \citealt{Hasinger08}, \citealt{Brusa10}), possibly linked to radiative and mechanical feedback, sweeping away the obscuring gas and dust. We do not obtain a significant correlation between these two observables (see Fig.~\ref{fig4}). However, the luminosity range ($43\leq \rmn{log}L_{2-10\,\rmn{keV}}\leq45$) is likely too narrow and the sample size too small to find such relation. 

Fig.~\ref{fig5} reports the best-fitting $N_H$ plotted against the adopted redshift for each source. We note that the majority ($\approx68$ per cent) of the sources are highly obscured ($N_H>10^{23}\rmn{cm^{-2}}$). Furthermore, there is a hint for an increasing trend of column density with redshift, as found in other works \citep[e.g. ][]{Treister06}. However, a possible bias due to the high redshift and low counting statistics can be present and will be investigated in the next section. The trend could be driven by the three sources at $z>5$, which have the most uncertain redshift information (see \S~\ref{sec2}). If they were placed at $z=3.7$ (the median redshift of the sample), the best-fitting $N_H$ would be $50^{+23}_{-18}\times10^{22}$, $42^{+20}_{-24}\times10^{22}$ and $118^{+41}_{-28}\times10^{22}\rmn{cm^{-2}}$ for XID 139, 197 and 485, respectively (i.e., they would still be in the highly obscured class). Assuming the photometric redshift ($z=4.42$) from \citet{Santini09} for XID 485, the only 
source in the sample at $z>5$ for which more than one photometric redshift is provided by the catalogues that we considered, the best-fitting column density would be $145^{+52}_{-35}\times10^{22}\rmn{cm^{-2}}$ (Fig.~\ref{fig5}, open circle). We recall that the low photon counting statistics allow us to assume only a simple spectral model, which does not take into account complex absorption structures \citep[e.g.][]{Bianchi12}, which could affect the results on the column density.

The four sources with $N_H>10^{24}\,\rmn{cm^{-2}}$ were also  fitted with the XSPEC model \textit{plcabs} (power-law emission transmitted through a spherical distribution of cold and dense matter, taking into account Compton scattering; \citealt{Yaqoob97}), which reproduces X-ray emission from Compton-thick sources better than a simple absorbed power-law, but no significant change in the best-fitting spectral parameters was found.

We also investigated the possible presence of a neutral iron K${\alpha}$ line in the spectra by adding a narrow ($\sigma=0.01$ keV) Gaussian component (\textit{zgauss} model in XSPEC) at 6.4 keV rest-frame. We could only find upper limits to the rest-frame equivalent width (EW) for most of the sources, of the order of a few 100 eV on average. This result is easily explained given that the observed equivalent width scales with $(1 + z)^{-1}$ and therefore only loose constraints can be placed to the line in spectra of high-redshift AGN (especially if the signal-to-noise ratio and the number of net counts are low and the accuracy in the redshift measurement is limited). The only source for which the iron line was easily visible in the spectrum and statistically significant is XID 412 (i.e. CDF-S 202 in \citealt{Norman02}). In this case, we allowed the rest-frame energy of the line to vary during the spectral fitting in a relatively large range. The best-fitting rest-frame energy, EW and $N_H$ that we derived 
for this source ($\rmn{E}=6.53$ keV, $\rmn{EW}=738^{+465}_{-451}$ eV, $N_H=91_{-13}^{+14}\times 10^{22}\rmn{cm^{-2}}$) are in good agreement with those found by previous \textit{Chandra} \citep[]{Norman02} and \textit{XMM-Newton} \citep[]{Comastri11} observations. The soft ($F_{0.5-2\,\rmn{keV}}=1.76\times10^{-16}\,\rmn{erg\,cm^{-2}s^{-1}}$) and hard ($F_{2-10\,\rmn{keV}}=2.41\times10^{-15}\,\rmn{erg\,cm^{-2}s^{-1}}$) fluxes are consistent with those reported by \cite{Tozzi06} using the 1 Ms CDF-S data, although they assumed a different spectral model ($\Gamma=1.8$ and $N_H=1.5\times10^{24}\rmn{cm^{-2}}$).

\subsection{$N_H$ distribution}
\label{nh}
\begin{figure}
\includegraphics[width=80mm,keepaspectratio]{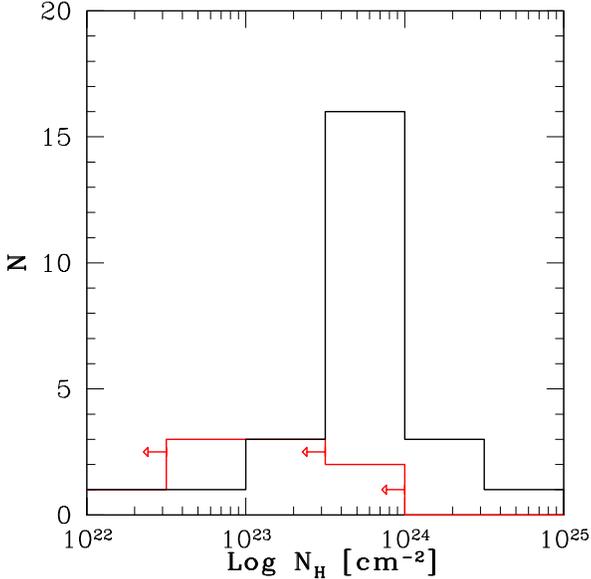}
\caption{Observed distribution of the best-fitting $N_H$. The black line represents the constrained values; the red line with leftward pointing arrows represents the upper limits.  }
\label{fig6}
\end{figure}

The distribution of the best-fitting $N_H$ (Fig.~\ref{fig6}) derived from spectral analysis, assuming an absorbed power-law with fixed $\Gamma=1.8$ as fitting spectral model, strongly suggests the presence of an important fraction ($\approx68\%$) of highly obscured sources ($N_H>10^{23}\rmn{cm^{-2}}$). This result could be affected by a systematic $N_H$ overestimate due to the high redshift and low counting statistics, characteristic of the sample: the photoelectric cut-off in spectra of mildly-obscured ($10^{22}<N_H<10^{23}\rmn{cm^{-2}}$) sources is redshifted out of the \textit{Chandra} band-pass already at $z=3-4$, making very difficult the search for the best-fitting $N_H$. Indeed, this type of spectrum might be indistinguishable from a spectrum of an unabsorbed source at high redshift. Furthermore, statistical fluctuations, enhanced by the low fluxes, are expected in the spectra and could simulate a spurious photoelectric cut-off at high rest-frame energies \citep[]{Tozzi06}.   

In order to estimate the probability ($P_{ij}$) that the X-ray spectrum of a source with intrinsic $N_H$ in a given range $j$ is fitted with a best-fitting column density constrained in a range $i$, we made extensive use of simulations. As in Tab.~\ref{tab1}, we define $N_H$ to be constrained if the $90\%$ c.l. lower limit on the best-fitting value is larger than zero. We divided the sample in 5 $N_H$ bins, as reported in Tab.~\ref{tab2}. The number of sources $N_i$ observed in each absorption bin $i$ is then given by:

\begin{equation}
 N_i = \sum_{j}x_jP_{ij}
\label{eq1}
\end{equation}

\noindent where $x_j$ is the number of sources with intrinsic $N_H$ in bin $j$ and $N_i$ is the number of sources observed with best-fitting $N_H$ in bin $i$, for $i$, $j$ = A, B, C, D, E. In this framework, $N_i$ is known (black histogram in Fig.~\ref{fig6}) and $P_{ij}$ can be derived from simulations. 

We simulated 1000 spectra with $\Gamma=1.8$ and 100 net counts (close to the median value of our sample), intrinsically obscured by a column density in the $j$-th  $N_H$ bin (for each bin we used the value reported in the third column of Tab.~\ref{tab2}), considering sources at $z=4$ (similar to the characteristic redshift of our sample), and simply counted how many times the simulated spectra are fitted with a best-fitting $N_H$ constrained in the $i$-th bin (see Appendix A for the detailed description of the simulation procedure). Bins A, B and C are indistinguishable at $z>3$ and hence are merged into a single bin (representing unobscured or mildly-obscured sources with $N_H<10^{23}\rmn{cm^{-2}}$), hereafter 
referred to as (ABC). 

 The value of each $P_{ij}$ is shown in Tab.~\ref{tab3}. We immediately note that no spurious effect seems to be important in bins D and E (i.e. spectra of intrinsically strongly obscured sources are efficiently fitted by a model obscured by a column density within the correct $N_H$ bin) while, as expected, the best-fitting $N_H$ is overestimated in most cases ($\approx64\%$) when the intrinsic value is in the bin (ABC).

\begin{table} 
\caption{$N_H$ bins used in simulations}
\label{tab2}
\begin{tabular}{ccc }
\hline
  \multicolumn{1}{c}{$i$,$j$} &
  \multicolumn{1}{c}{range: $\rmn{log}(\frac{N_H}{\rmn{cm^{-2}}})$}&
  \multicolumn{1}{c}{sim: $\rmn{log}(\frac{N_H}{\rmn{cm^{-2}}})$}\\
\hline
A & $<21$ & 0.0 \\[1ex]
B & 21-22 & 21.5\\[1ex]
C & 22-23 & 22.5\\[1ex]
D & 23-24 & 23.5\\[1ex]
E & $>$24 & 24.5\\[1ex]
\hline                                    

\end{tabular}
\medskip
\\Sources in the $i$-th or $j$-th $N_H$ bin (first column), which accounts for the $N_H$ range shown in the second column, are simulated assuming the column density reported in the third column.
\end{table}

\begin{table} 
\caption{Final probability factors $P_{ij}$ that a source with intrinsic $N_H$ in a given bin $j$ is observed in a bin $i$. These are derived from simulations using $\Gamma=1.8$ and $\Gamma=1.6$.}
\label{tab3}
\begin{tabular}{cccc }
\hline
  \multicolumn{1}{c}{$P_{ij}$} &
  \multicolumn{1}{c}{$j$=(ABC)}&
  \multicolumn{1}{c}{$j$=D}&
  \multicolumn{1}{c}{$j$=E}\\
\hline
\hline
\multicolumn{4}{c}{$\Gamma=1.8$}\\[1ex]
\hline
$i$=(ABC)& 0.361 & 0.0 & 0.0\\[1ex]
$i$=D    & 0.639 & 1.0 & 0.0\\[1ex]
$i$=E    & 0.0   & 0.0 & 1.0\\[1ex]
\hline     
\hline
\multicolumn{4}{c}{$\Gamma=1.6$}\\[1ex]
\hline
$i$=(ABC)& 0.147 & 0.0 & 0.0\\[1ex]
$i$=D    & 0.853 & 1.0 & 0.0\\[1ex]
$i$=E    & 0.0   & 0.0 & 1.0\\[1ex]
\hline 
\hline
\end{tabular}
\medskip

\end{table}

\begin{figure}
\includegraphics[width=80mm,keepaspectratio]{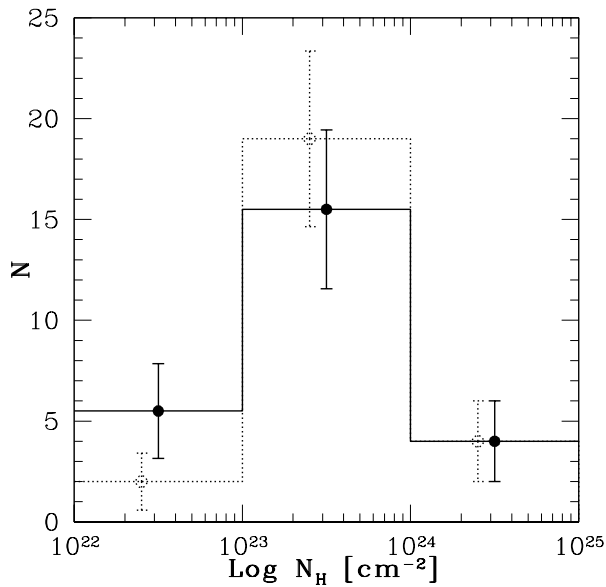}
\caption{Intrinsic $N_H$ distribution (solid line) with statistical errors, derived correcting the observed one (dotted line, plotted in the same bins, with statistical errors slightly shifted for visual purpose), resulting from spectral analysis assuming $\Gamma=1.8$, with the probability factors obtained from simulations (see \S~\ref{nh} and Appendix \ref{AppendixA}). Only constrained values of $N_H$ are considered.}
\label{fig7}
\end{figure}

\begin{figure}
\includegraphics[width=80mm,keepaspectratio]{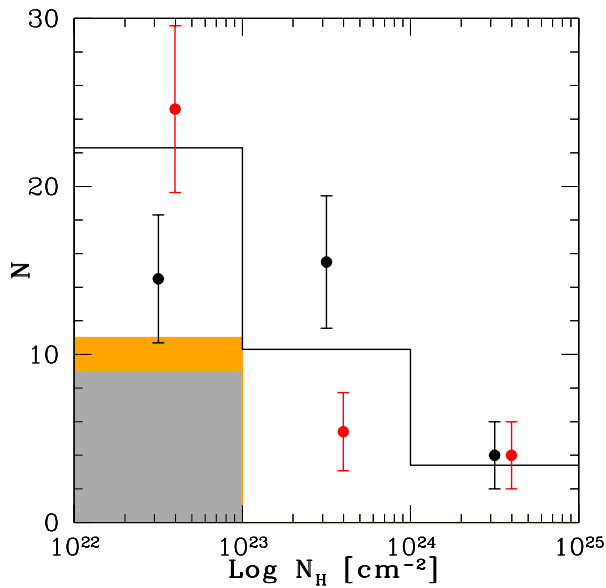}
\caption{Intrinsic $N_H$ distribution assuming $\Gamma=1.8$ (black points) and $\Gamma=1.6$ (red points, which are slightly shifted from the centre of the bins for visual purposes). The first bin includes sources with $N_H<10^{23}\rmn{cm^{-2}}$ (i.e. consistent with no or ``moderate" absorption) and all the sources for which only upper limits on $N_H$ were obtained (9 and 11 sources in the two cases, represented as grey and orange shaded areas, respectively). The third bin contains sources with $N_H>10^{24}\rmn{cm^{-2}}$. Poissonian errors are computed for the number of sources in each bin. The histogram is the predictions computed with the POMPA tool and corrected for the sky coverage of the survey (see text). The predictions are not normalized to the observed number of sources.}
\label{fig8}
\end{figure}

By inverting eq. (\ref{eq1}), the intrinsic $N_H$ distribution can now be derived (Fig.~\ref{fig7}): $x_\rmn{(ABC)}\approx5.5$, $x_\rmn{D}\approx15.5$ and   $x_\rmn{E}=4$, to be compared with the observed distribution: $N_{\rmn{(ABC)}}=2$, $N_{\rmn{D}}=19$ and $N_{\rmn{E}}=4$. If we conservatively count all the 9 sources for which only upper limits could be obtained (red line in Fig.~\ref{fig6}) in the first bin, the resulting $N_H$ distribution is the one shown in Fig.~\ref{fig8}. We added the predictions of the \citet*[hereafter GCH07]{Gilli07} X-ray background (XRB) synthesis model computed with the POMPA\footnote{POrtable Multi-Purpose Application for AGN counts \\(http://www.bo.astro.it/$\sim$gilli/counts.html)} tool for each $N_H$ bin, considering sources with luminosity $42  \leq \rmn{Log}\frac{L_{0.5-2\,\rmn{keV}}}{\,\rmn{erg}\,\rmn{s^{-1}}}\leq 47$ in a redshift range between $z_{min}=3$ and $z_{max}=8$. A high-redshift exponential decline in the space density of AGN, similar to that of bright 
quasars \citep[]{Gilli11}, is also considered. The assumption of the decline affects only the normalization and not the shape of the distribution. The predictions are corrected for the sky coverage  (i.e. the sky area in physical units covered by the survey at different fluxes, see next section), to take into account the smaller sky area covered by the survey at low fluxes. Obscured sources have a flux lower than unobscured or less obscured ones with the same luminosity at the same redshift and, hence, can be detected over a smaller, deeper area.

The $N_H$ distribution, corrected for the spurious overestimate of the column density, shows a number of strongly obscured sources that is slightly larger than the predictions of the model we considered. Conversely, the number of unobscured or mildly-obscured sources is smaller than the predicted number. However, we find an agreement between the observations and the predictions at the $2\sigma$ confidence level (the $\chi^2$ test returns a value of 6.0 for 3 degrees of freedom and no free parameter).

Following the results of \S~\ref{sec4} where, considering an absorbed power-law with $\Gamma$ free to vary as starting spectral model, we derived a mean photon index $<\Gamma>\approx1.6$, we investigated the effects of assuming a flatter photon index on the column density distribution. Therefore, we fitted again the X-ray spectra of all sources in the sample using the same spectral model with photon index fixed to $\Gamma=1.6$. The resulting $N_H$ distribution was corrected for spurious overestimates using the probability factors (Tab.~\ref{tab3}) obtained from simulations, similarly to the $\Gamma=1.8$ case, but now using $\Gamma=1.6$ to simulate and fit the spectra. Red points in Fig.~\ref{fig8} represent the intrinsic $N_H$ distribution in the $\Gamma=1.6$ case. We derived upper limits on $N_H$ for two sources (XID 132 and 700, which were previously counted in bin D), while their best-fitting column density was constrained using $\Gamma=1.8$. This result produces a different number of sources in the two 
shaded histograms in Fig.~\ref{fig8}. Therefore, the number of objects in each bin of the corrected $N_H$ distribution is, in this case, $x_\rmn{(ABC)}\approx13.6$, $x_\rmn{D}\approx5.4$ and $x_\rmn{E}=4$. If we place all the 11 sources with an upper limit on $N_H$ in the first bin, as in Fig.~\ref{fig8}, the distribution is fully consistent within $2\sigma$ with the prediction of the model ($\chi^2=4.75$ with 3 degrees of freedom and no free parameter). We underline that the model predictions are computed with the same level of incompleteness as the observed sample.

\subsection{The logN-logS of the high-redshift sample}
We derived the logN-logS of the high-redshift sample in the soft band, where the survey sensitivity is the highest, folding the number $N$ of sources with flux larger than $S$ with the survey sky-coverage at that flux. We assumed the soft fluxes reported in Tab.~\ref{tab1} derived from our spectral analysis. The sky-coverage was obtained by multiplying the nominal sky area covered by the survey ($\approx0.13\,\rmn{deg^2}$; \citealt{Xue11}) and the fraction of the f.o.v. covered at different fluxes, derived applying the CIAO tool DMIMGHIST on the sensitivity map \citep[]{Xue11}, which assigns to each point of the f.o.v. the flux limit of the survey at that position. In order to be consistent with the computation of the sky coverage (derived in the $0.5-2\,\rmn{keV}$ band), we excluded four ($\approx12\%$ of the total sample) sources which were not detected in the soft band (see Tab.~\ref{tab1}). 

\begin{figure}

 \includegraphics[width=80mm,keepaspectratio]{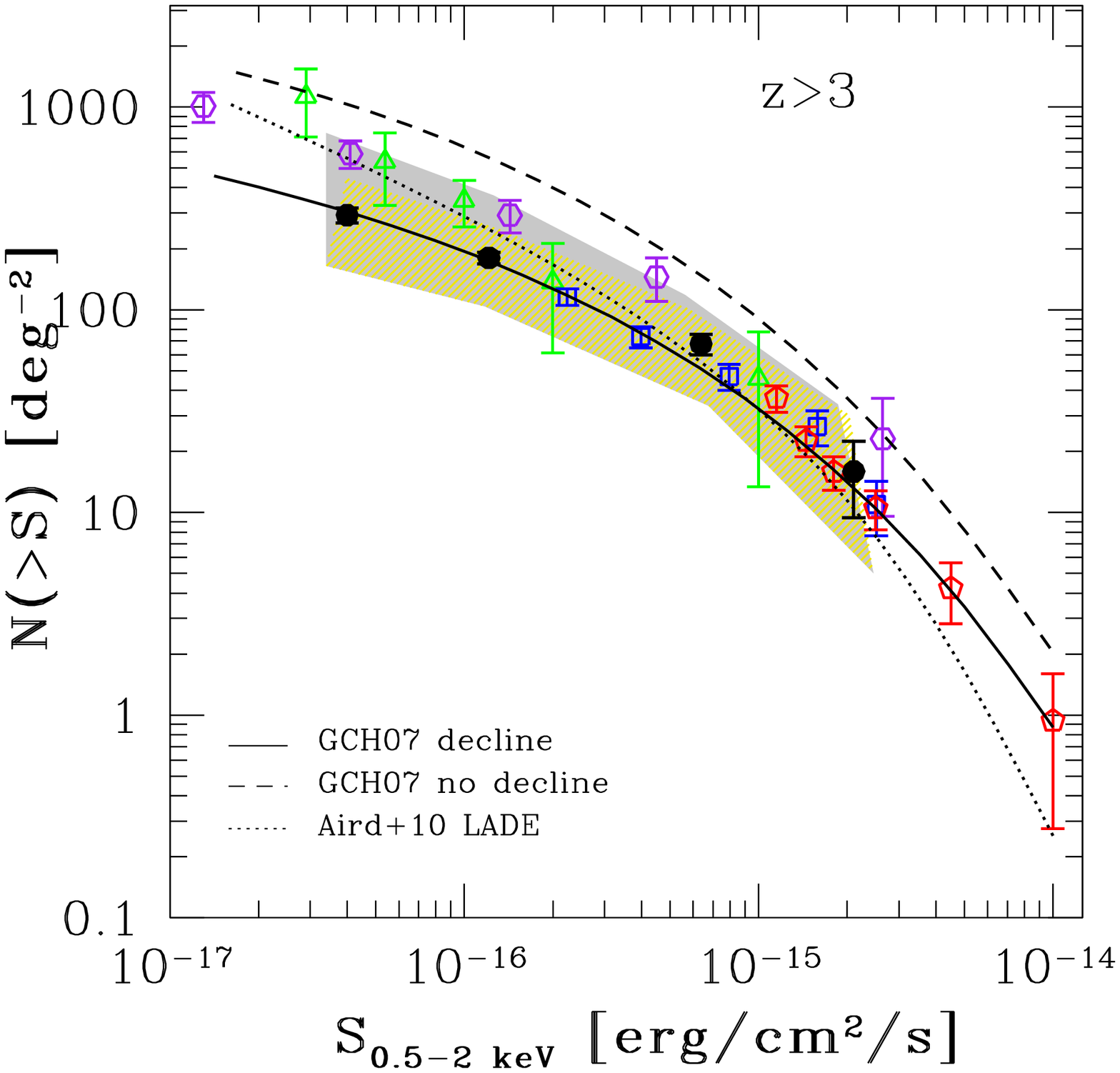}
 \includegraphics[width=80mm,keepaspectratio]{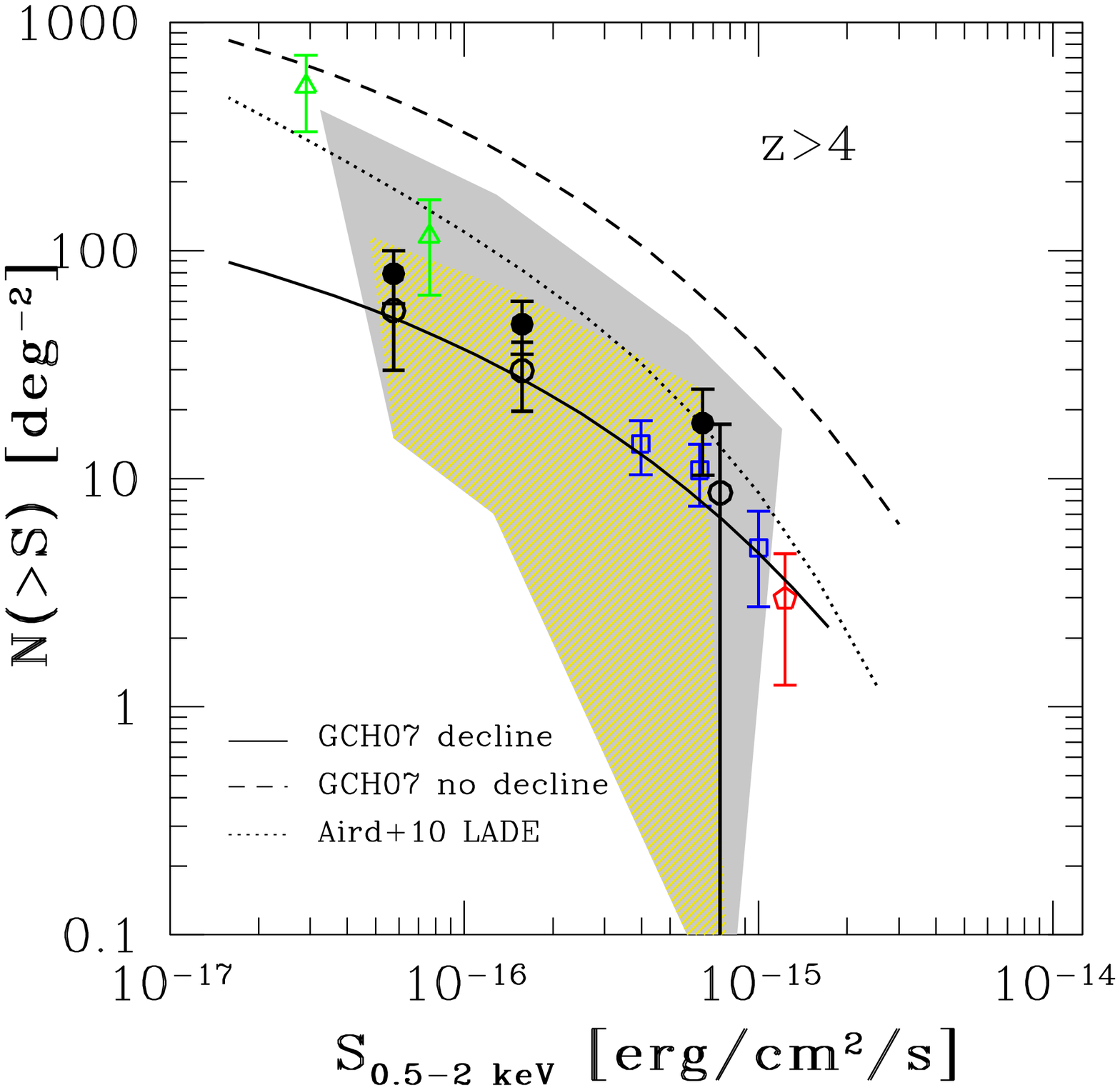}
\caption{Binned LogN-logS of the high-redshift sample, considering all sources ($z>3$, \textit{top panel}) and only those at $z>4$ (\textit{bottom panel}), with associated Poissonian errors. Black filled circles are from this work, green triangles are from \citet{Fiore12}, blue squares are from \citet{Civano11}, red pentagons are from \citet{Brusa09} and purple hexagons are from \citet{Lehmer12}. Empty black circles in the bottom panel represent our data if the three sources at the highest redshifts were placed at $3<z<4$ (see text). The predictions of the \citeauthor{Gilli07} XRB synthesis model, computed with the POMPA tool, either assuming (solid line) or not (dashed line) an exponential decline in the AGN space density at high redshift, and \citet{Aird10} model are also shown. The yellow shaded area represents the regions in which the LogN-LogS would be placed if more conservative or shallower selection criteria were used as well as the grey area, which also includes the 39 sources with no redshift in 
the 
considered catalogues (see \S~\ref{discuss}).} 

\label{fig9}

\end{figure}

In Fig.~\ref{fig9} we compare our data with those provided by \citet{Fiore12} and \citet{Lehmer12} for the 4 Ms CDF-S  and with the XMM \cite[]{Brusa09} and \textit{Chandra} \cite[]{Civano11} COSMOS data. The predictions of the \citeauthor{Gilli07} XRB synthesis model (solid and dashed lines) and the \cite{Aird10} X-ray luminosity function (XLF) evolutionary model (dotted line) are also shown. The \citeauthor{Gilli07} model is based on a luminosity dependent density evolution (LDDE) model of the XLF with redshift, and it allows the presence of a high-redshift ($z>2.7$) decline in the AGN space density to be taken into account. The \cite{Aird10} model assumes a luminosity and density evolution (LADE) of the XLF.

Our results for the $z>3$ sample (\textit{top panel}) are in good agreement with the \citeauthor{Gilli07} model predictions assuming an exponential decline in high-redshift AGN space density (continuous line), down to flux $F_{0.5-2\,\rmn{keV}}\approx4\times10^{-17}\rmn{erg}\,\rmn{cm^{-2}s^{-1}}$, a factor $\approx10$ fainter with respect to previous surveys (i.e. down to $L_X\approx10^{43}\,\rmn{erg}\,\rmn{s^{-1}}$, the minimum luminosity detectable at $z>3$ in the 4 Ms CDF-S): we extend to lower fluxes the results obtained in the COSMOS field by \citet{Brusa09} and \citet{Civano11}, confirming an evolution at high redshift similar to that of the more luminous AGN \citep{Gilli11}.

Considering only the $z>4$ subsample (Fig.~\ref{fig9}, \textit{bottom panel}, filled circles), our data lie above the model predictions. We include here the presence of 3 sources at $5<z\la7.7$, whose redshifts are determined on the basis of relatively uncertain photometric information. If these sources were placed at $3<z<4$, a good agreement would be obtained (Fig.~\ref{fig9}, \textit{bottom panel}, open circles).

The scenario without the decline (Fig.~\ref{fig9}, dashed lines) is excluded by our data both at $z>3$ and $z>4$.

\section{Discussion}\label{discuss}
 
The $N_H$ distribution derived from spectral analysis, corrected for the overestimate of the column density, suggests the presence of a significant fraction ($\sim57$ per cent) of strongly obscured ($N_H>10^{23}\rmn{cm^{-2}}$) AGN at $z>3$. This result was obtained by fitting the X-ray spectra with an absorbed power-law with $\Gamma=1.8$, i.e. the typical photon index observed in large AGN samples \citep[e.g.][]{Turner97, Tozzi06}. Adopting $\Gamma=1.6$, we derived a slightly different intrinsic $N_H$ distribution, with more unobscured sources, with respect to the previous case. Since the current data are sampling a rest-frame energy range up to at least $E_{int}\approx20\,\rmn{keV}$, a hardening of the X-ray spectra due to reflection might be expected at observed energies $E_{obs}\approx4-5\,\rmn{keV}$. However, the low photon-counting statistics prevented us from using relatively complex spectral models and the $\Gamma=1.6$ index may simply incorporate the hardening due to reflection. Therefore, the ``true"
 $N_H$ distribution is expected to lie between the two distributions derived assuming $\Gamma=1.8$ and $\Gamma=1.6$ (Fig.~\ref{fig8}). To further investigate the possible impact of a reflection component, we computed the stacked, rest frame, X-ray spectra  (using the method described in \citealt{Iwasawa12}) dividing the sample in four $N_H$  bins ($\frac{N_H}{10^{23}\,\rmn{cm^{-2}}} < 1.3$; $1.3<\frac{N_H}{10^{23}\,\rmn{cm^{-2}}}<6$; $6<\frac{N_H}{10^{23}\,\rmn{cm^{-2}}}<9$; $\frac{N_H}{10^{23}\,\rmn{cm^{-2}}}>9$; see table 1). The resulting best fit slope, obtained by a joint fit of the the various $N_H$ bins and representing the intrinsic continuum above 10 keV, is $\Gamma\simeq 1.69^{+0.07}_{-0.07}$. This slope is suggestive of a limited contribution from reflection and is well within the values adopted to compute the absorption distribution (Fig. 8). Independently from the assumed spectrum, the data are in agreement at the $2\sigma$ confidence level with the prediction of the \citeauthor{Gilli07} XRB 
synthesis model, assuming an exponential decline in the AGN space density at $z>2.7$ (Fig.~\ref{fig8}). The best-fitting column densities of the 19 high-luminosity ($L_{2-10\,\rmn{keV}}>10^{44}\rmn{erg}\,\rmn{s^{-1}}$) sources are well constrained at high values ($\sim84$ per cent of them has nominal $N_H>10^{23}\rmn{cm^{-2}}$). Applying the correction factors derived assuming $\Gamma=1.8$ (see \S~\ref{nh}), we derived that the intrinsic fraction of obscured sources ($N_H>10^{23}\rmn{cm^{-2}}$) in this subsample is $\sim66$ per cent.

The \citeauthor{Gilli07} model assumes that the fraction of obscured AGN does not evolve with redshift. A possible evolutionary scenario emerging from recent works suggests that this behaviour could be valid only for quasars (QSOs) with $L_{2-10\,\rmn{keV}}\le10^{44}\rmn{erg}\,\rmn{s^{-1}}$ (see \citealt{Gilli10}), which is the median luminosity of our sample. Indeed, evidences are found that the activity of luminous ($L_{2-10\,\rmn{keV}}\ge10^{44}\rmn{erg}\,\rmn{s^{-1}}$) AGN, triggered by major gas-rich mergers, occurs in short bursts \citep[with timescales of the order of $\sim0.01\,\rmn{Gyr}$; e.g.][]{Alexander05}. In this case, the gas accretion is chaotic \citep{Hopkins08} and may produce a large covering factor and column densities. Since the merging rate increases with redshift, the fraction of obscured AGN may then increase as well, as reported in e.g. \cite{LaFranca05}, \cite{Hasinger08} and Iwasawa et al. (2012, submitted). At lower luminosities, the accretion onto the SMBH is probably 
driven by secular processes \citep[$\sim1\,\rmn{Gyr}$; e.g.][]{Daddi07, Elbaz11} and the accretion is expected to be smoother and symmetrical. Therefore, the fraction of obscured AGN would be quite constant with redshift, given that obscuration would be a function of the system geometry only. Our results fit well in this framework.

Fig.~\ref{fig9} shows the number counts derived in this work compared to previous determinations from the literature and theoretical models. The CDF-S source counts  down to $F_{0.5-2\,\rmn{keV}}\approx4\times10^{-17}\rmn{erg}\,\rmn{cm^{-2}s^{-1}}$ lie on the XRB model if an exponential decline of the AGN X-ray luminosity function is assumed, similarly to \textit{XMM} \citep[]{Brusa09} and \textit{Chandra} \citep[]{Civano11} COSMOS results at brighter fluxes. In particular, we note the agreement at bright fluxes with the results from COSMOS surveys, for which a major effort was made to obtain spectroscopic and photometric redshifts \citep[][and references therein]{Salvato09, Brusa09, Salvato11}. Our results are consistent also with the \citet{Aird10} model if the least conservative bounds (grey area in Fig.~\ref{fig9}) are assumed (see below for the computation of the error budget). At fluxes $2\times10^{-16} \le F_{0.5-2\,\rmn{keV}}\le 2\times10^{-15}\rmn{erg}\,\rmn{cm^{-2}s^{-1}}$ (i.e. for the brightest 
sources in 
the high-redshift sample) the prediction of this model is very similar to the \citeauthor{Gilli07} model, but at fainter fluxes the two models deviate significantly. Our results better trace the \citeauthor{Gilli07} model, while data from \citet{Fiore12} and \citet{Lehmer12} are placed on the \citet{Aird10} model.

There is consistency in the number counts that we present with data from \citet{Fiore12} and \citet{Lehmer12} down to $F_{0.5-2\,\rmn{keV}}\approx10^{-16}\rmn{erg}\,\rmn{cm^{-2}s^{-1}}$, while our results are a factor $\sim2-3$ lower at fainter fluxes. This is not surprising considering that different selection techniques are applied, ours being more conservative than the ones used in those works. Indeed, the logN-logS in \citet{Fiore12} is derived for 54 optically selected sources at $z>3$, using spectroscopic and photometric redshifts and drop-out techniques and applying a new detection algorithm on the 4 Ms CDF-S observations, which made possible to reach a flux limit of $F_{0.5-2\,\rmn{keV}}\approx1.2\times10^{-17}\rmn{erg}\,\rmn{cm^{-2}s^{-1}}$. The samples considered in \citet{Lehmer12} and in this work are collected from the same catalogue of X-ray selected sources. However, \citet{Lehmer12} assumed the redshift adopted in \cite{Xue11}, while we collected additional information from other 
catalogues and used a stricter selection procedure (as described in \S~\ref{sec2}). Therefore, we rejected a number of sources which are included in \citet{Lehmer12}, and we were able to collect redshift information for some objects which have no redshift in \cite{Xue11}, as in the case of XID 331. We would obtain results largely consistent with \citet{Lehmer12} using their sample.

 To compute the ``error budget" on the number counts, we investigate how many sources would be included in the sample if we relax the third, fourth and fifth criteria  described in \S~\ref{sec2}. Relaxing the third criterion and including in the final sample soft X-ray sources for which $N_{phot,>3} - N_{phot,<3} \geq 0$, the net gain in the final sample would be of 4 sources. As for the fourth criterion, 14 sources were excluded from the sample because only one of the two available photometric redshift is at $z>3$. We consider these objects to have a $50\%$ chance of being at $z>3$, and randomly include seven of them in the error budget.
Finally, if we relax the fifth criterion and include in the sample all sources with only one photometric redshift, with the condition of being at $z>3$, regardless of its error, we would add 4 sources to the final sample. If all these changes were adopted, the sample would consist of 49 sources.
 
 The gold shaded area in Fig.~\ref{fig9} indicates the regions in which the logN-logS would lie considering the error budget \footnote{We assigned the soft X-ray fluxes reported in \citet{Xue11} to the sources not included in the high-redshift sample for the computation of all shaded areas.}. For the $z>3$ number counts (upper panel), the lower bound is obtained if only sources selected via the first, second and third criteria of \S~\ref{sec2} are included in the high-redshift sample (i.e. sources with spectroscopic redshift or $N_{phot,>3} - N_{phot,<3} \geq 3$). This is clearly a very conservative assumption. The upper bound is obtained if we relax the third, fourth and fifth criteria, as discussed above, adding 15 sources in total. 

Four out of these 15 sources would be at $z>4$: two would be selected relaxing the fifth criterion and two relaxing the fourth criterion. As previously described, we randomly assumed half of the latter sources to be at $z>4$. The resulting 3 sources were added to the $z>4$ sample to compute the upper-bound of the gold shaded area (bottom panel in Fig.~\ref{fig9}), which accounts for the error budget of the $z>4$ number counts. The lower bound is obtained, as before, considering only the first, second and third criteria, and consists of one source with a spectroscopic redshift $z>4$ (XID 717, while we discarded XID 403 because it is not detected in the soft band) and 2 sources selected through the third criterion. As expected, because of the low number of sources detectable at $z>4$, only loose constraints can be placed on the LogN-LogS. Larger samples of X-ray selected AGN at $z>4$ would be needed to draw more solid conclusions.

Thirty-nine sources detected in the soft band in \citet{Xue11} have no redshift in any catalogue that we considered. A not negligible fraction of them are probably AGN at $z>3$. The grey shaded regions in Fig.~\ref{fig9} show how the gold area would enlarge if all these sources would be accounted for. Therefore, these areas represent the hypothetical and unlikely cases in which all the 39 sources were at $z>3$ ($z>4$).

\section{Conclusions}
 The detection of a significant number of $z>3$ AGN down to $L_X\approx10^{43}\,\rmn{erg}\,\rmn{s^{-1}}$ is made possible by the faint flux limit reached in the 4 Ms CDF-S observations \citep[]{Xue11}. Moreover, X-rays are less biased against obscuration with respect to the optical band. The large multi-wavelength coverage of the CDF-S allows us to retrieve spectroscopic and photometric redshifts for most of the sources in the field. In this framework, physical and evolutionary properties of a relatively large sample of high-redshift AGN can be studied.

 The sample consists of 34 sources in the range $3<z\le7.6$, with a median redshift of $z=3.7$. About 45 per cent of them (15 sources) have a spectroscopic redshift. 

X-ray spectra of all the sources were extracted from the 4 Ms CDF-S observations and fitted with an absorbed power-law spectral model with photon index fixed to $\Gamma=1.8$. The low photon-counting statistics prevents us from assuming a more complex spectral model: the median full-band ($0.5-7\,\rmn{keV}$) net-count number is 80. The median rest-frame absorption-corrected luminosity of the sample is $L_{2-10\,\rmn{keV}}\approx1.5\times10^{44}\rmn{erg}\,\rmn{s^{-1}}$.

We summarise the conclusions reached in this work as follows:

\begin{enumerate}
\item The observed column density distribution is strongly peaked between $23<log(\frac{N_H}{\rmn{cm^{-2}}})<24$. The low number of counts and the high redshift of the source sample may lead to overestimate the intrinsic absorption. To account for this effect, we ran extensive X-ray spectral simulations and derived correction factors to apply to the observed distribution. Therefore we obtained an estimate of the intrinsic $N_H$ distribution. The results are consistent with a non-evolving obscured fraction of AGN with $L_x\approx10^{44}\rmn{erg}\,\rmn{s^{-1}}$ with redshift within $2\sigma$.

\item The number counts of our $z>3$ AGN sample are consistent with a decline in the AGN space density at high redshift, extending at faint soft  fluxes ($F_{0.5-2\,\rmn{keV}}\approx4\times10^{-17}\rmn{erg}\,\rmn{cm^{-2}s^{-1}}$) the behaviour determined at brighter fluxes found by \citet{Brusa09} and \citet{Civano11} in the COSMOS field. We used a more conservative approach with respect to \cite{Fiore12} and \cite{Lehmer12}: if we significantly relax our selection criteria, our results are in a relatively good agreement at bright fluxes with their data, while at the faint end the discrepancy is a factor $\sim2$.

These results would be improved by increasing the size and the reliability of the sample (e.g. gathering additional reliable redshifts via ultradeep optical-IR spectroscopic campaigns) and the quality of the X-ray spectra (e.g. increasing the \textit{Chandra} exposure in this field). The accuracy of the photometric redshifts would be significantly improved by adding the deep Cosmic Assembly Near-IR Deep Legacy Survey (CANDELS) H and J-band data \citep[]{Grogin11, Koekemoer11}.

\end{enumerate}

\section*{Acknowledgments}
We acknowledge J. Aird for providing the LADE model data, M. Mignoli for his help with optical spectra, E. Vanzella for providing some spectroscopic redshifts and T. Dahlen for providing his catalogue of photometric redshifts. We thank G. Hasinger for the helpful suggestions which improved this work. We acknowledge financial support from the agreement ASI-INAF I/009/10/0 and the PRIN-INAF-2011. WNB and BL acknowledge financial support CXC grant SP1-12007A and NASA ADP grant NNX10AC99G. YQX thanks the Youth 1000 Plan (QingNianQianRen) program and the USTC startup funding (ZC9850290195). FEB acknowledges support from Basal-CATA (PFB-06/2007), CONICYT-Chile (FONDECYT 1101024), and {\it Chandra} X-ray Center grant SAO SP1-12007B. This research has made use of data obtained from the Chandra Data Archive and software provided by the Chandra X-ray Center (CXC) in the application packages CIAO.

\bsp

\appendix

\section{Deriving the correction factors for the observed $N_H$ distribution from simulations of X-ray spectra}\label{AppendixA}
The correction factors discussed in \S~\ref{nh} are derived from spectral-fitting simulations using XSPEC, assuming five different degrees of obscuration (see \S~\ref{nh} and Tab.~\ref{tab2}). The five cases are simulated by runs of 1000 simulations each. The spectra are simulated at $z=4$ with an input normalization tuned to obtain a distribution of net counts peaked at $\approx100$ (with a resulting dispersion of $\sigma\approx11\%$), using the response files of a real source at an off-axis angle $\theta=4\,\rmn{arcmin}$. The results do not vary significantly if different response files are used, but they are sensitive to the number of net counts and redshift. To be consistent with the real cases, for these parameters we assumed values close to the average ones of the sample.

The simulation procedure can be summarized in the following steps:
\begin{enumerate}
 \item Definition of an input model: we considered a power-law with photon index fixed to $\Gamma=1.8$ as input model, absorbed by five different column densities, whose values vary between the five considered cases (Tab.~\ref{tab2}, third column). Galactic absorption ($N_H=7\times10^{19}\rmn{cm^{-2}}$) is also included. The desired number of counts is obtained by properly adjusting the normalization of the power-law.
\item  Simulation of the spectrum: the \textit{fakeit} command in XSPEC can simulate a spectrum with background, given a starting model. It requires as input the response and ancillary files, a real background file (with the POISSERR keyword set to TRUE) and the desired exposure time. We set it to $3.5\times10^6\rmn{s}$ to be consistent with the average real case. The output consists of a fake source and a fake background spectrum file. The same BACKSCAL keyword, read from the header of the input background file, is assigned by the \textit{fakeit} procedure to both simulated spectra, which are therefore assumed to be extracted from regions with an equal area. This approach is not correct and causes the source and background counts to be wrongly scaled (i.e. a strong underestimate of the source net-count rate). To address this problem, before simulating the spectra we multiplied the EXPOSURE keyword of the real background file by a factor equal to the ratio between the BACKSCAL parameters of the real 
background and 
associated source 
file. We checked that the errors on the net-count rate were correctly computed (i.e. propagating the errors of the total and background-only count-rates).

\item Spectral fitting: after having grouped the simulated spectra, using the FTOOL \textit{grppha}, with  at least 1 count per bin, and assigned to them the response files and the fake background spectra, the spectra are repeatedly fit in the energy range $\rmn{E}=0.5-5\,\rmn{keV}$ with an absorbed power-law model with $N_H$ free to vary and photon index fixed to $\Gamma=1.8$, accounting also for the Galactic absorption (fixed to $N_H=7\times10^{19}\rmn{cm^{-2}}$). Cash statistics were employed in the fitting. The fitting step (which use a local minimization algorithm) is alternated to the computation of the errors of the various free parameters (which is a more global operation) for a few times, in order to avoid the fit to be stuck in a local minimum.
\end{enumerate}

For each of the five considered cases of input $N_H$, we obtained as output 1000 best-fitting values of $N_H$. Following the procedure adopted to fit the real spectra, we define $N_H$ to be constrained if the lower limit (at the 90 per cent c.l.) on the best-fitting value is larger than zero, otherwise we consider its upper limit (e.g. if the fit returns a best-fitting value $N_H=10^{+12}_{-10}\times10^{22}\rmn{cm^{-2}}$, we consider $N_H<22\times10^{22}\rmn{cm^{-2}}$). 

The probability $p_{ij}$ that a simulated source in the $j$-th $N_H$ bin is observed in the $i$-th $N_H$ bin, are derived by counting the number of times a constrained best-fitting $N_H$ falls in bin $i$ and normalized this number to 1000. In order to conserve the number of sources when the correction factors are applied on the real distribution, the $p_{ij}$ factors must be rescaled to the total probability to constrain $N_H$ (i.e. we excluded all cases in which only upper limits could be found):
\begin{equation}
 P_{ij}=\frac{p_{ij}}{\sum_{i=A}^{E}p_{ij}}
\end{equation}
As discussed in \S~\ref{nh}, bins A, B and C are merged into a single bin (ABC), since they are indistinguishable at $z>3$. Therefore, the probability factors, $P_{i\rmn{(ABC)}}$, are:
\begin{equation}
 P_{\rmn{(ABC)(ABC)}}=\sum_i\frac{\sum_jp_{ij}}{3}
\end{equation}

for $i$, $j$ = A, B, C; $P_{i\rmn{(ABC)}}=\frac{p_{i\rmn{A}}+p_{i\rmn{B}}+p_{i\rmn{C}}}{3}$ and $P_{\rmn{(ABC)}j}=\frac{p_{\rmn{A}j}+p_{\rmn{B}j}+p_{\rmn{C}j}}{3}$ for $i$, $j$ = D and E.

The same procedure is then repeated using $\Gamma=1.6$ instead of $\Gamma=1.8$.
The $P_{ij}$ factors, for $i$ and $j=$ (ABC), D and E,  are those used in \S~\ref{nh}.

\section{X-ray spectra of the $\lowercase{z}>3$ sample}

 \begin{figure}
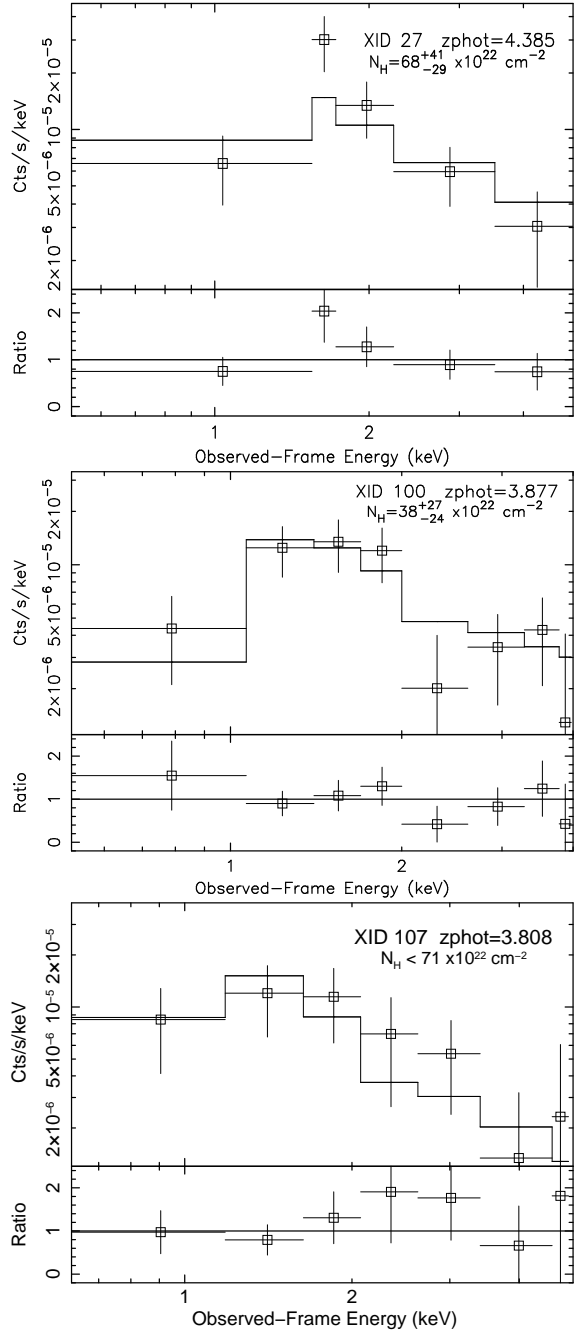

\begin{center}
 \includegraphics[width=62mm,height=75mm,angle=270]{27_spec.ps}\\
  \includegraphics[width=57mm,height=75mm,angle=270]{100_spec.ps}\\
\includegraphics[width=57mm,height=75mm,angle=270]{107_spec.ps}
\end{center}
\caption{X-ray spectra of the 34 sources in the high-redshift sample, fitted with an absorbed power-law model with $\Gamma=1.8$ (see \S~\ref{sec4}). The ratio of data to folded model is shown and the spectrum is binned (at $\ge3\sigma$ detection significance) for display purposes only. Identification numbers are from \citet{Xue11}. The spectroscopic or photometric redshift is reported for each source, as well as the best-fitting column density,  accordingly to Tab.~\ref{tab1}. (Continue)}
\end{figure}
\addtocounter{figure}{-1}

 \begin{figure*}
\begin{center}
\includegraphics[width=57mm,height=75mm,angle=270]{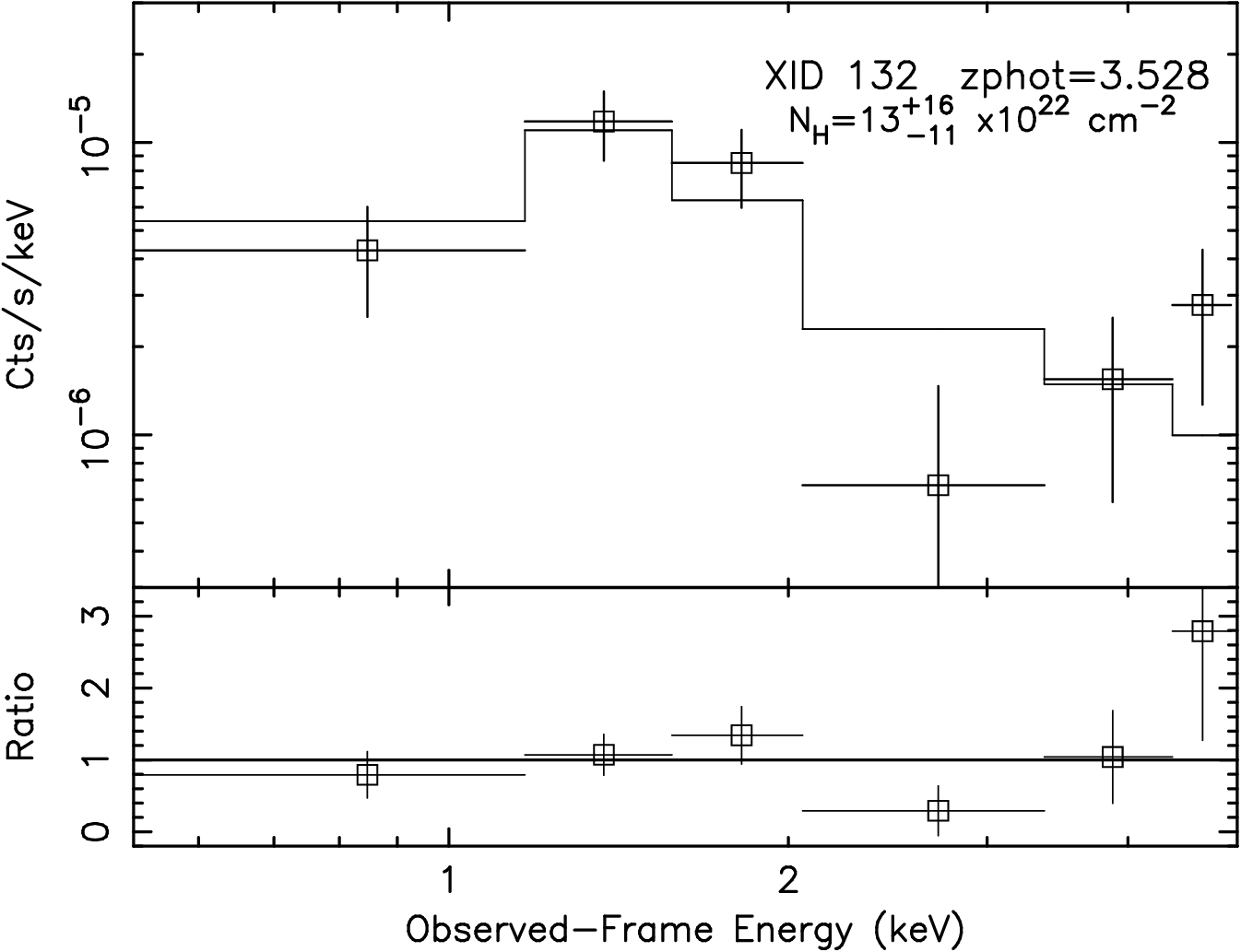}
\includegraphics[width=57mm,height=75mm,angle=270]{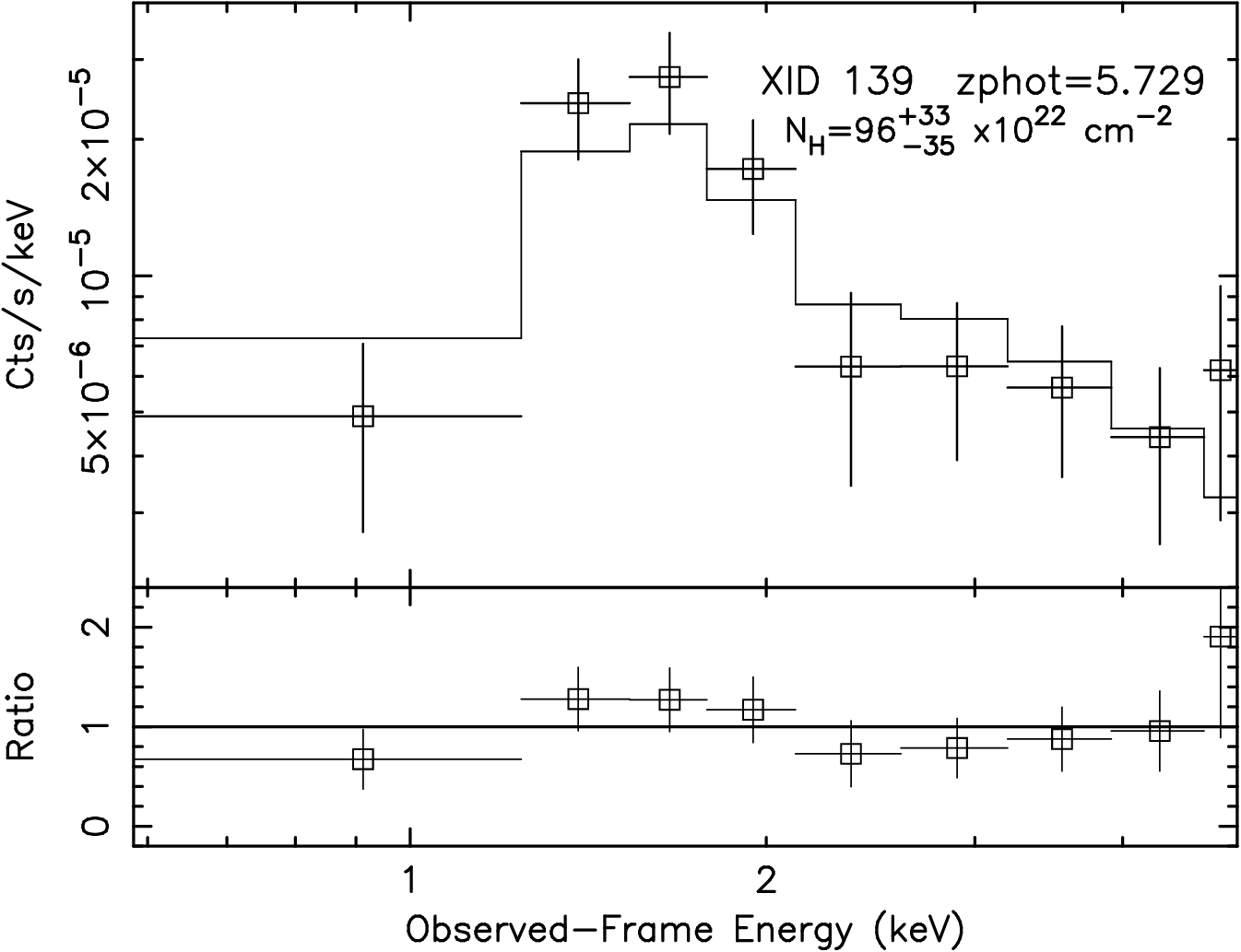}
\includegraphics[width=57mm,height=75mm,angle=270]{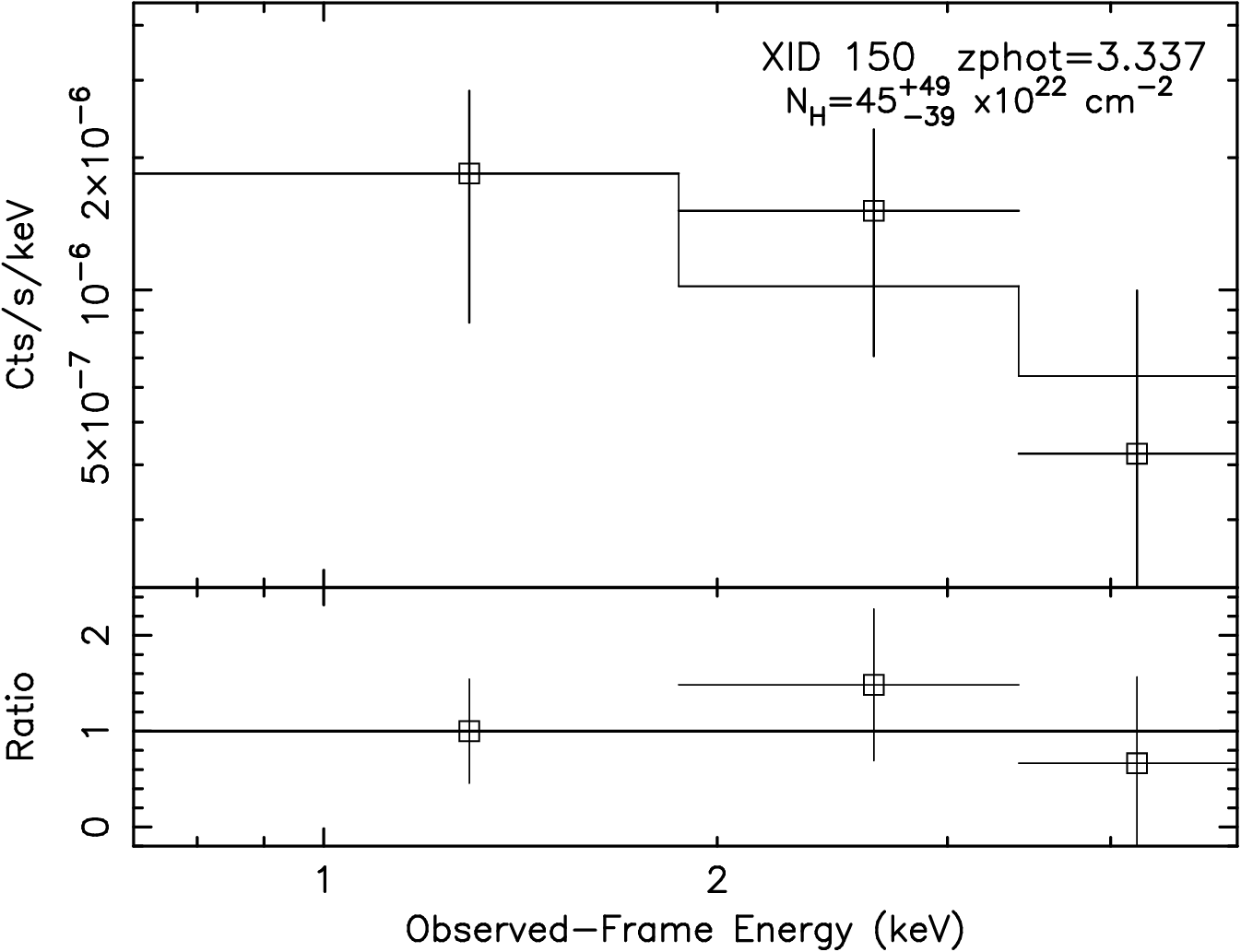}
\includegraphics[width=57mm,height=75mm,angle=270]{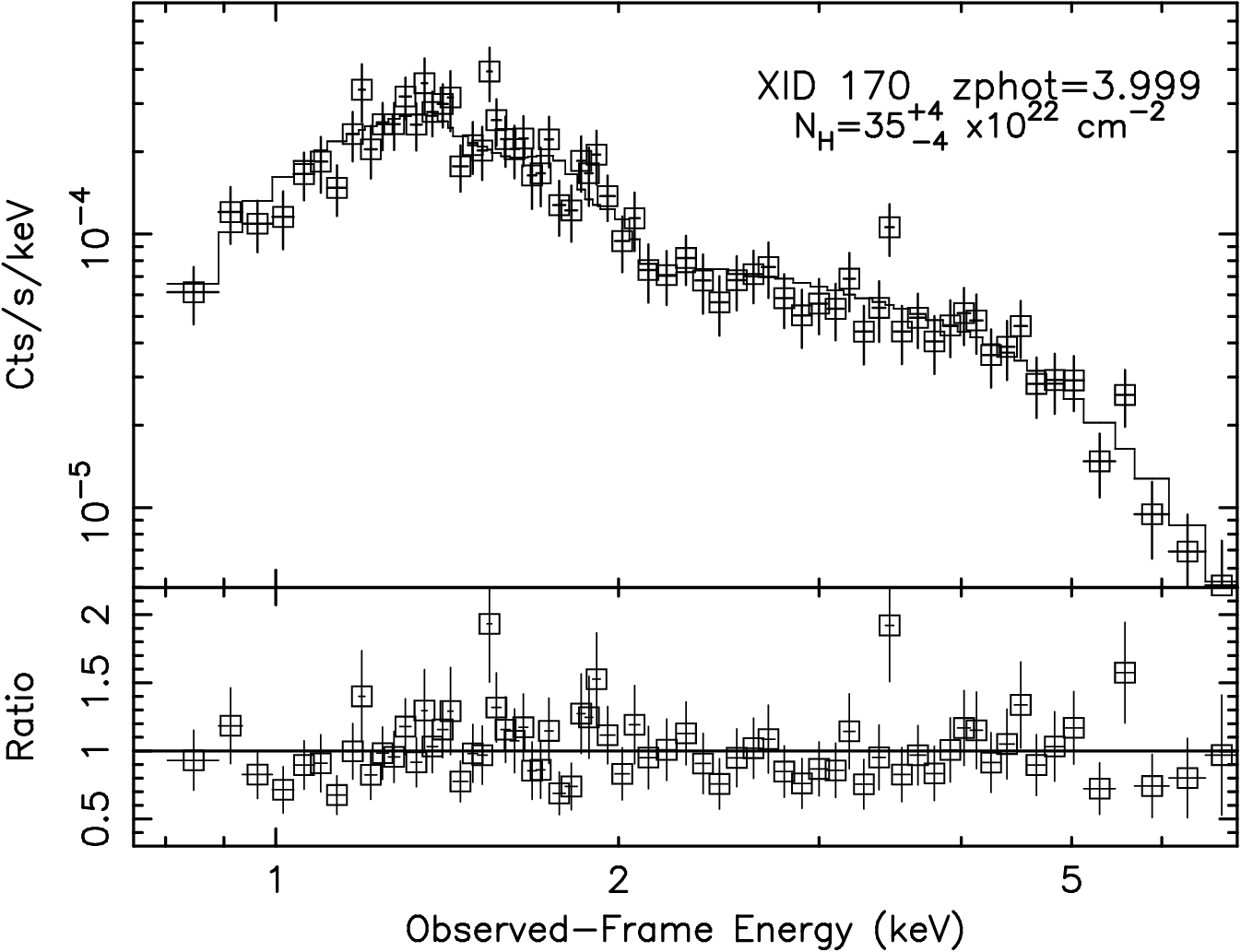}
\includegraphics[width=57mm,height=75mm,angle=270]{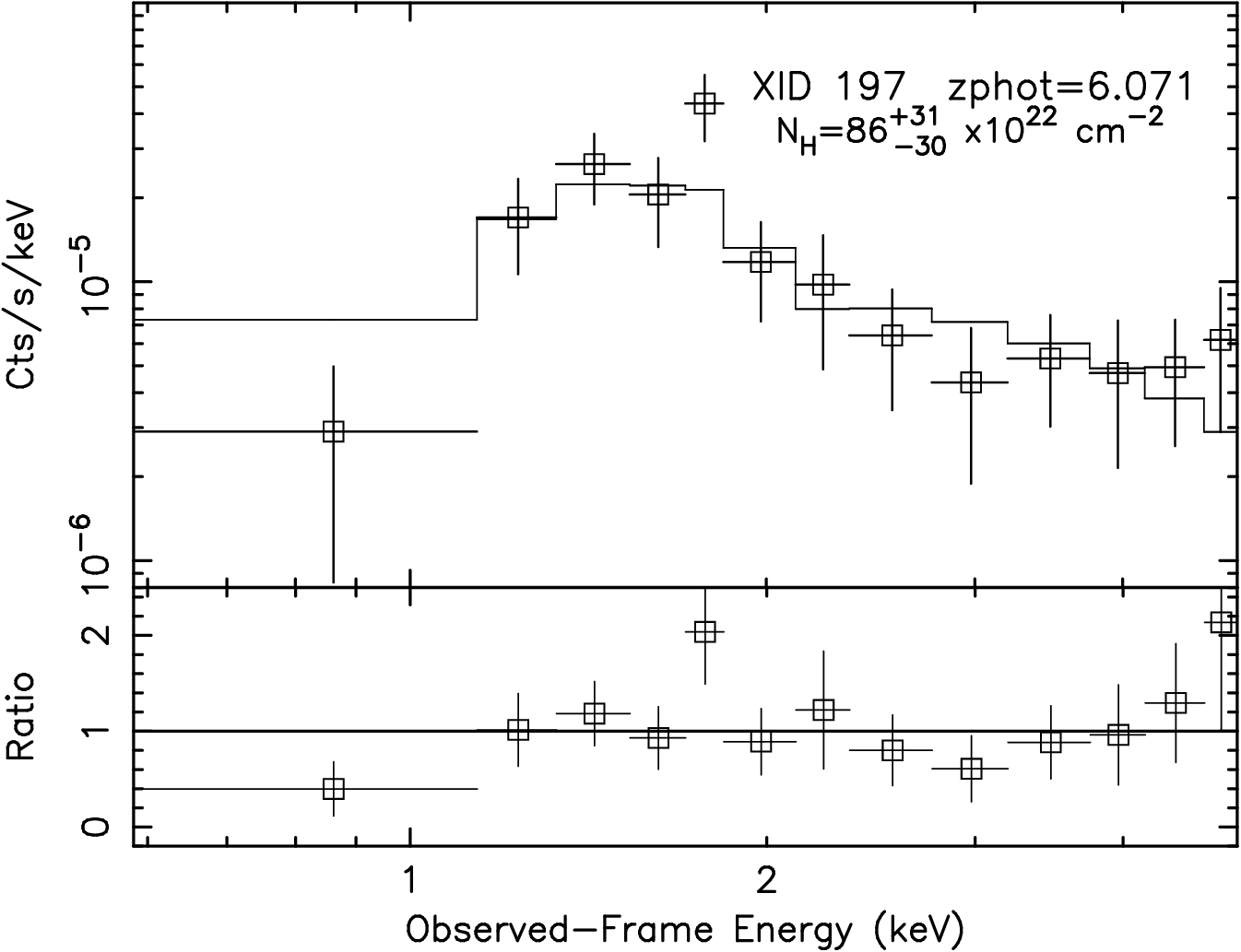}
 \includegraphics[width=57mm,height=75mm,angle=270]{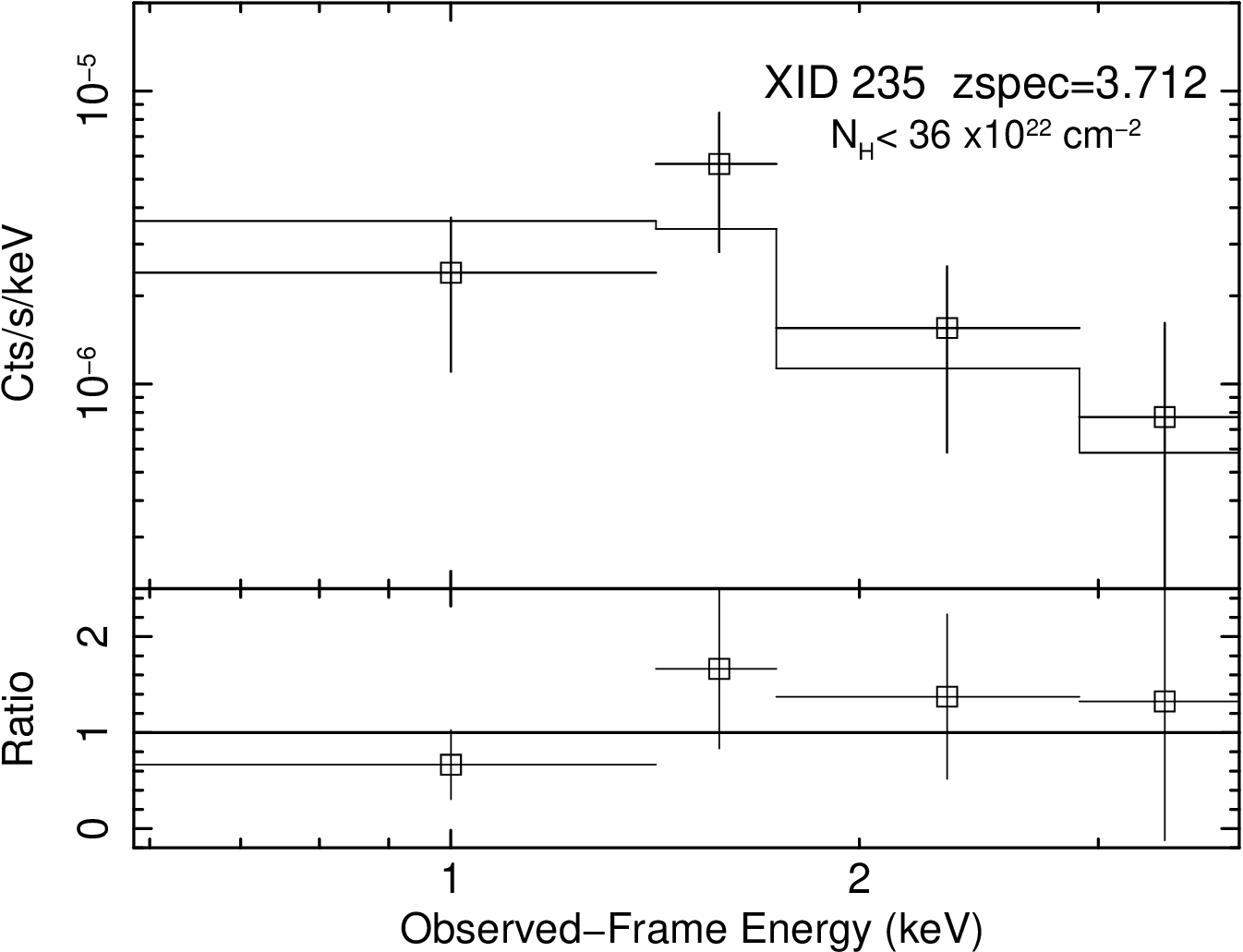}
 \includegraphics[width=57mm,height=75mm,angle=270]{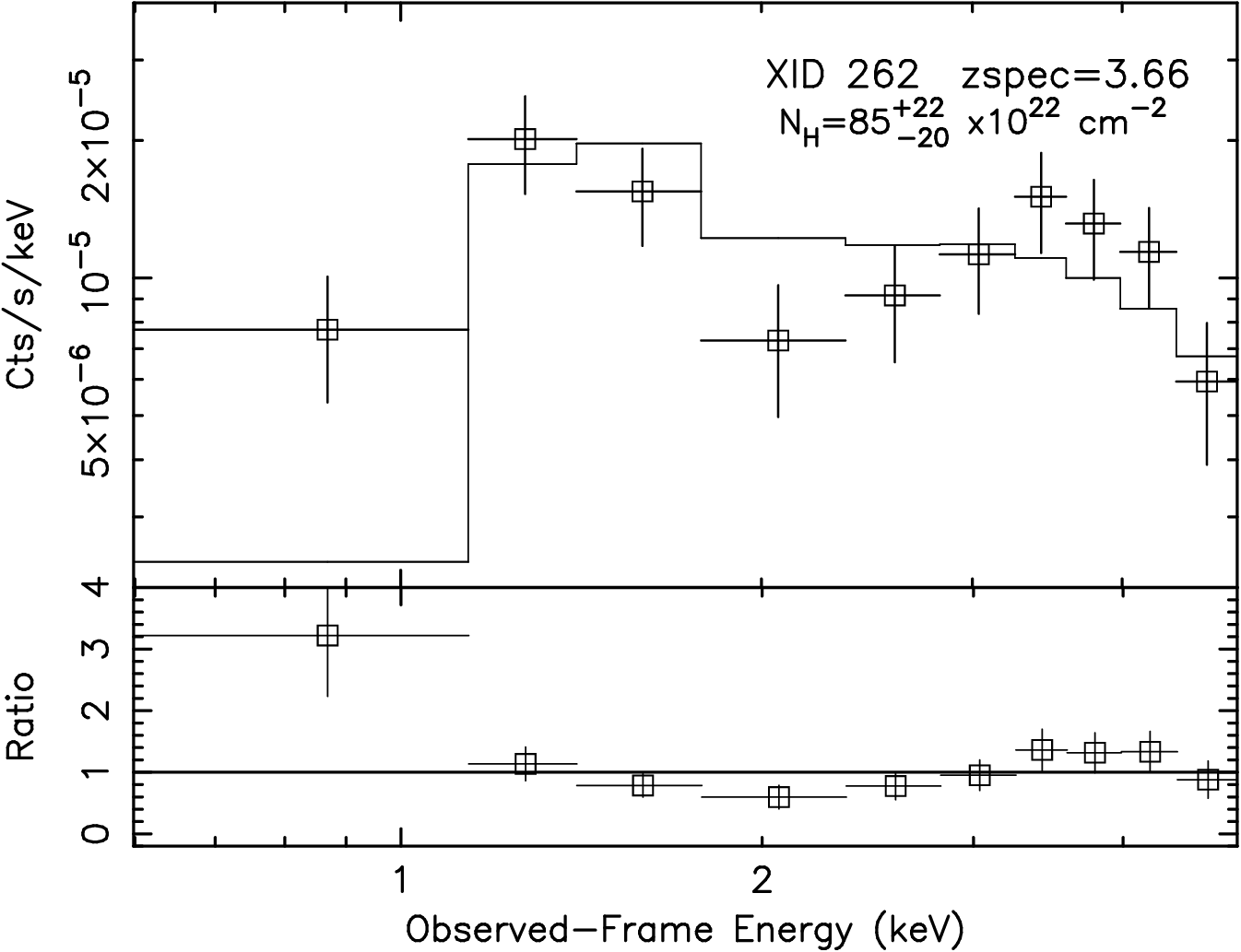}
 \includegraphics[width=57mm,height=75mm,angle=270]{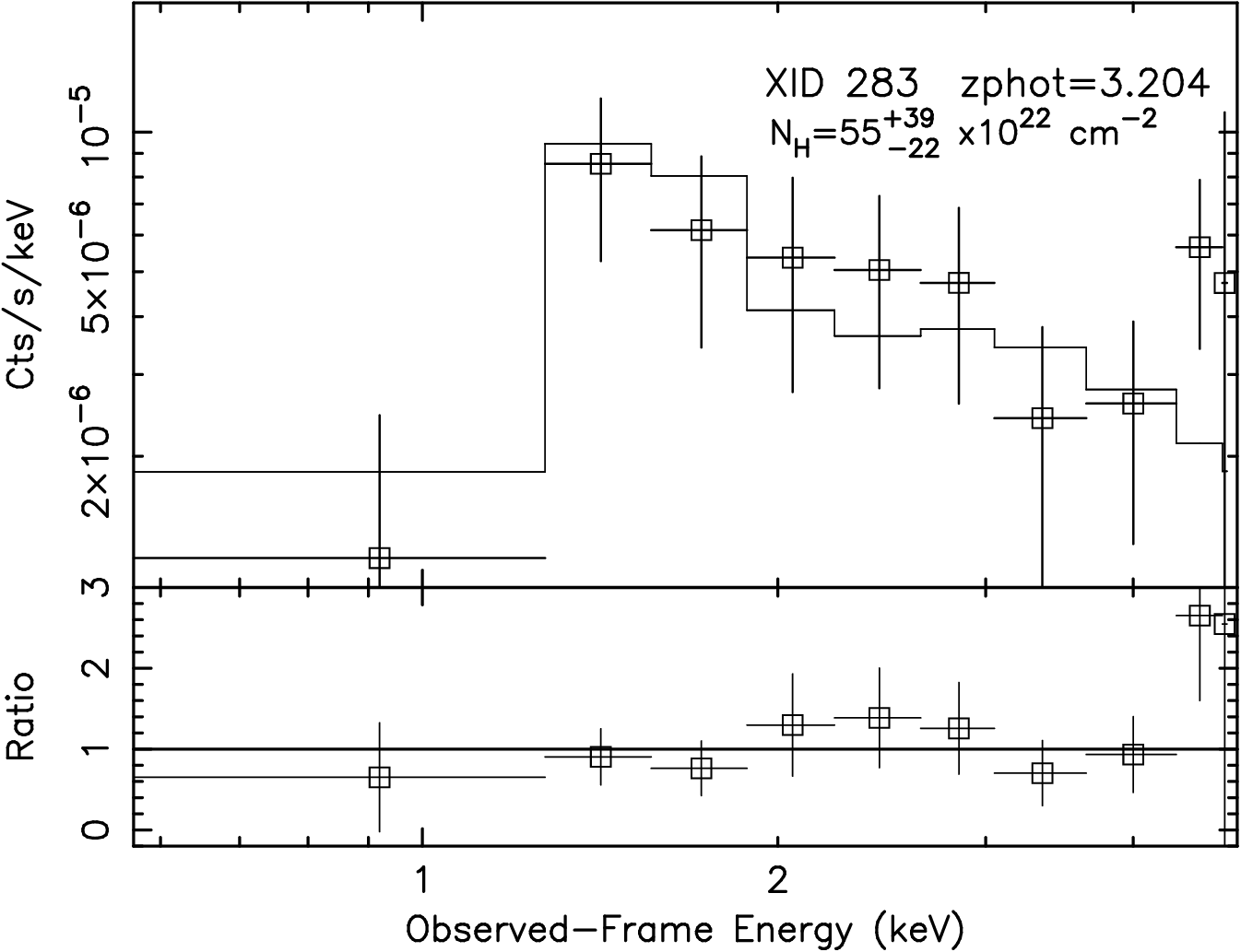}
\end{center}
\caption{(Continue)}
\end{figure*}
\addtocounter{figure}{-1}
\begin{figure*}
\begin{center}

 \includegraphics[width=57mm,height=75mm,angle=270]{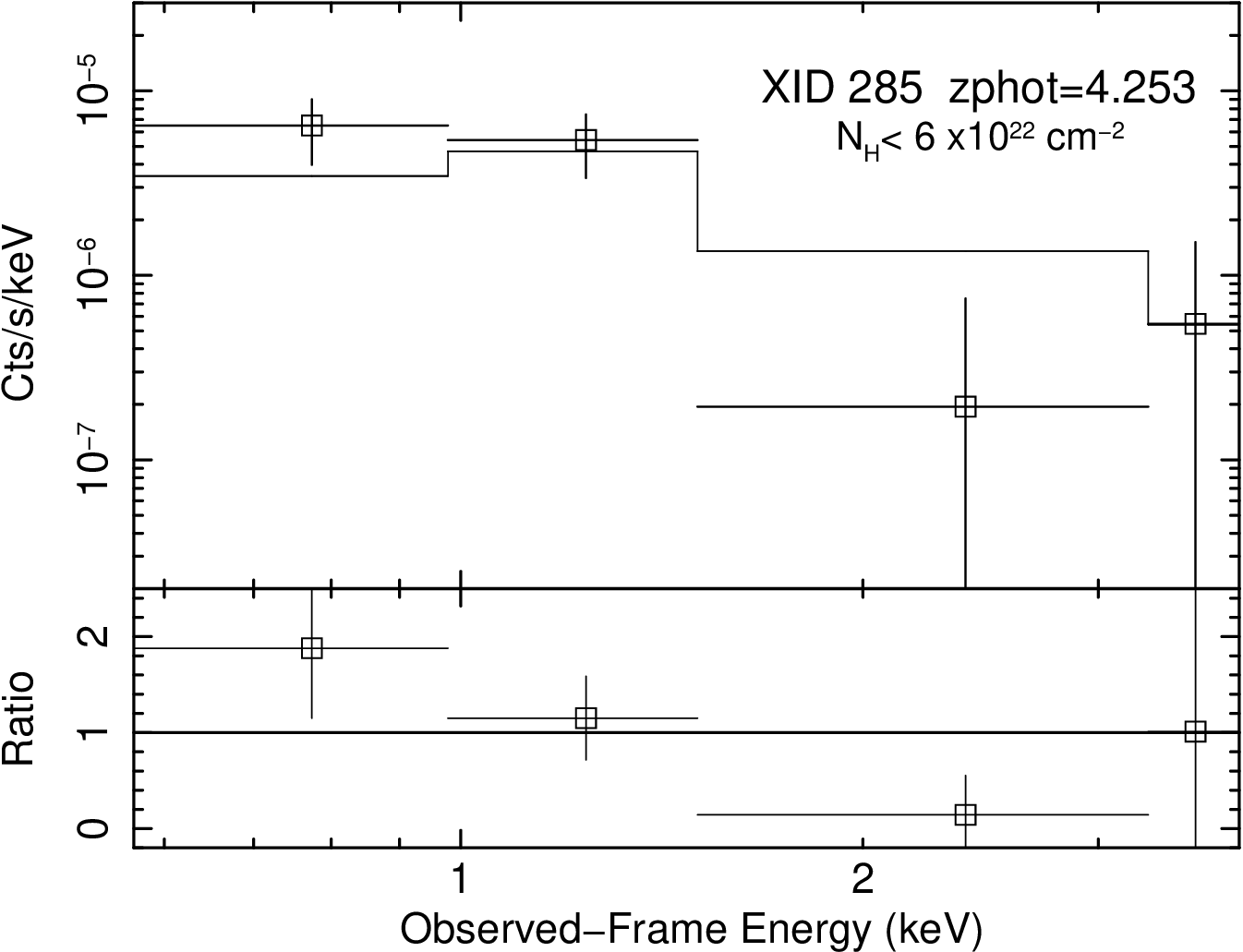}
 \includegraphics[width=57mm,height=75mm,angle=270]{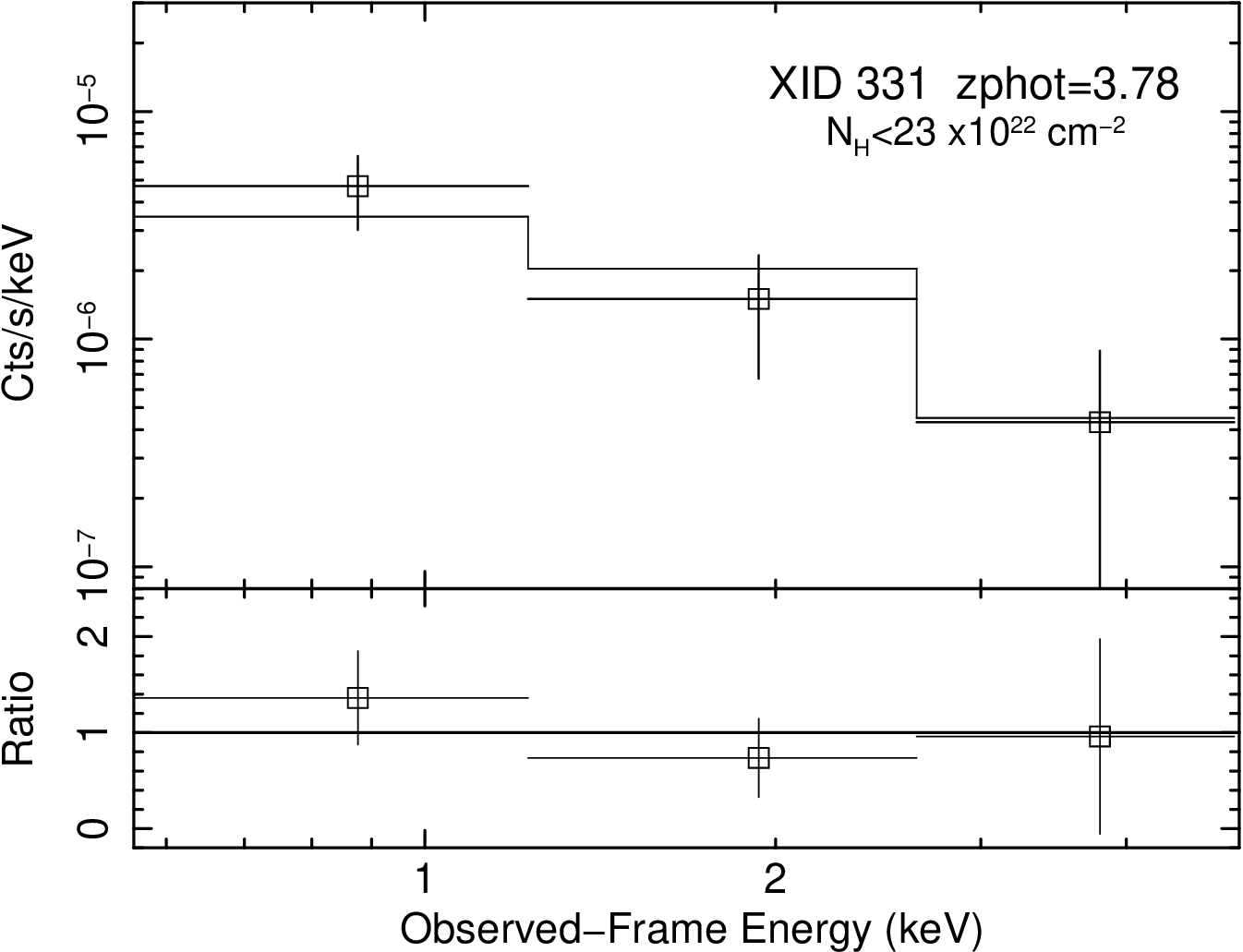}
 \includegraphics[width=57mm,height=75mm,angle=270]{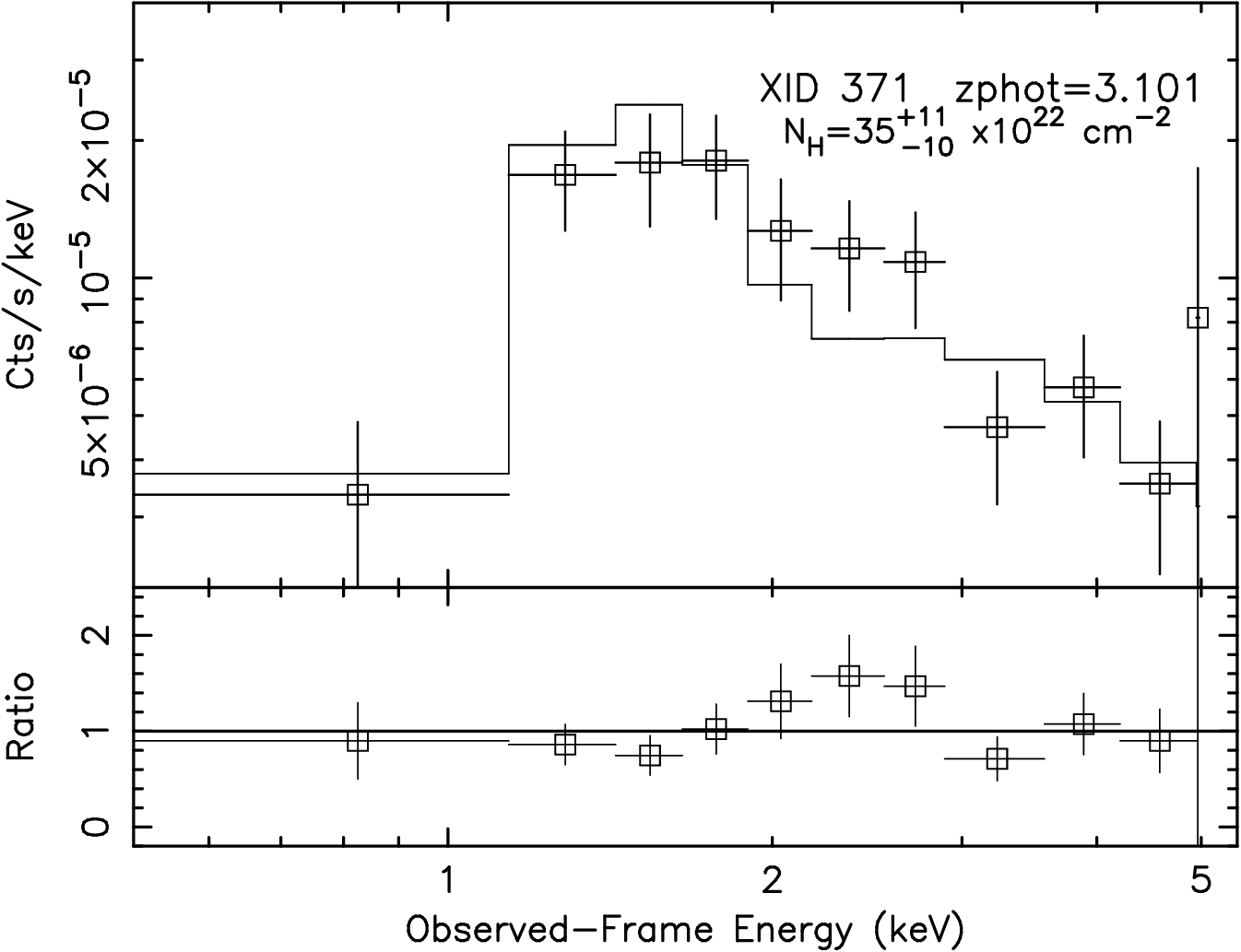}
  \includegraphics[width=57mm,height=75mm,angle=270]{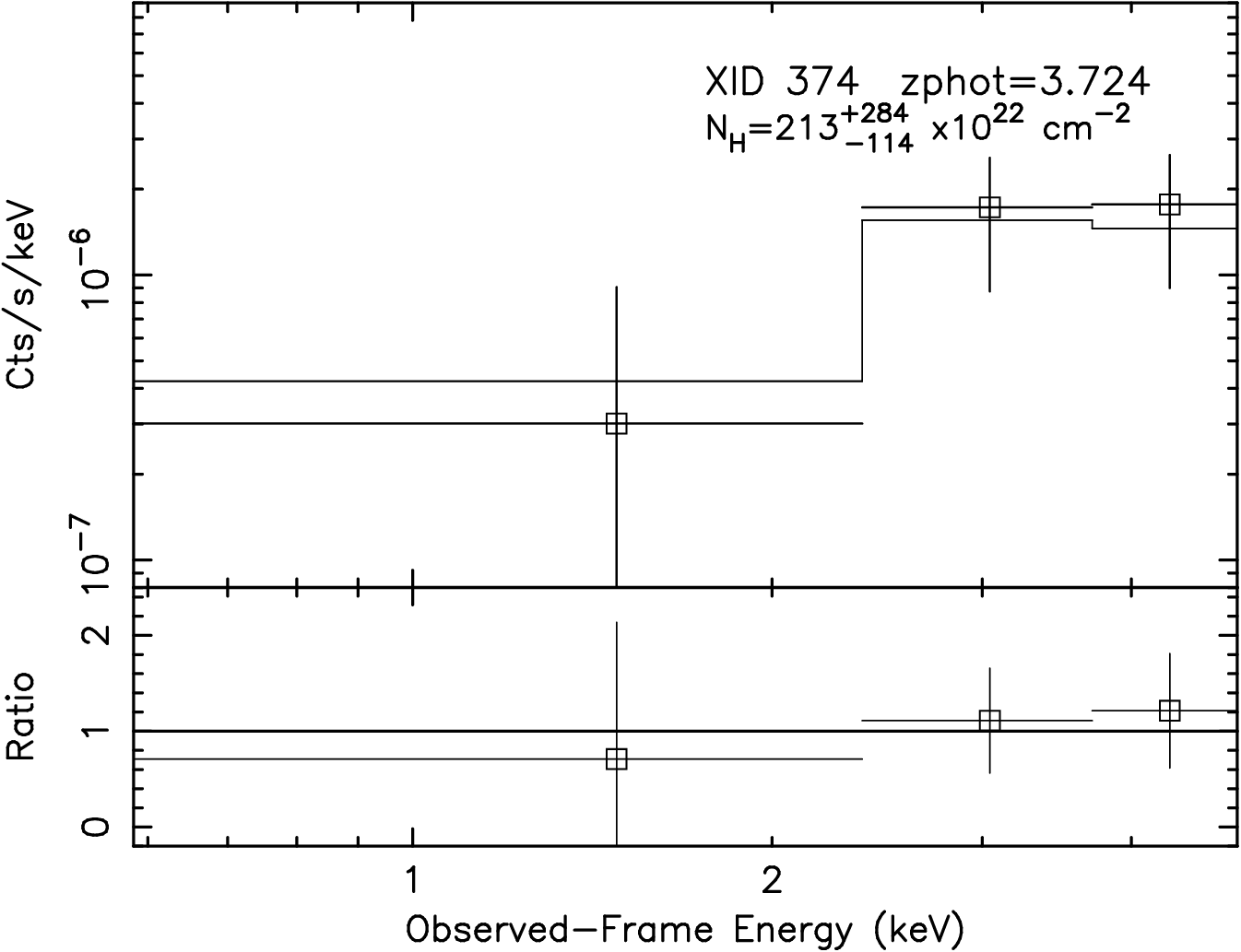}
 \includegraphics[width=57mm,height=75mm,angle=270]{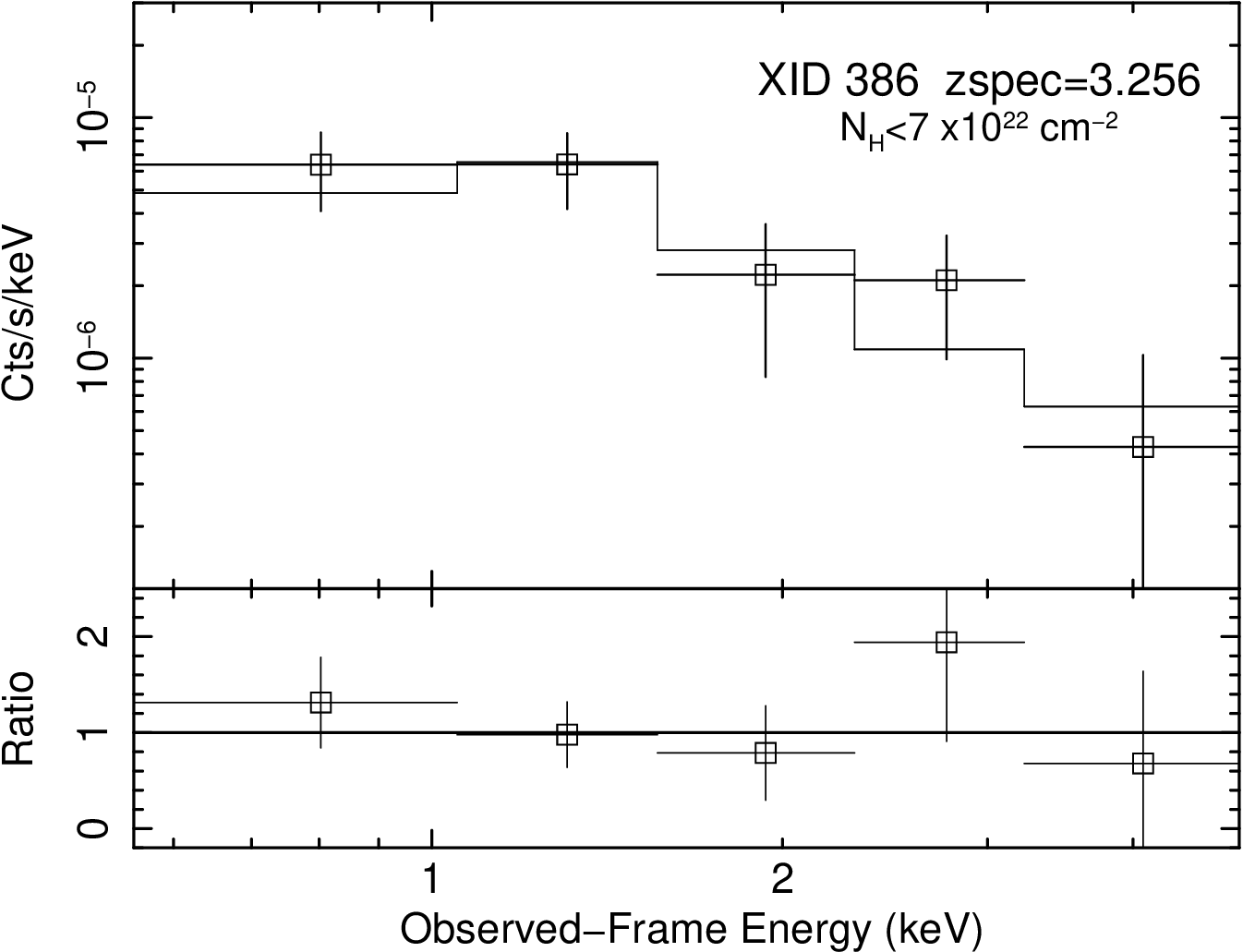}
 \includegraphics[width=57mm,height=75mm,angle=270]{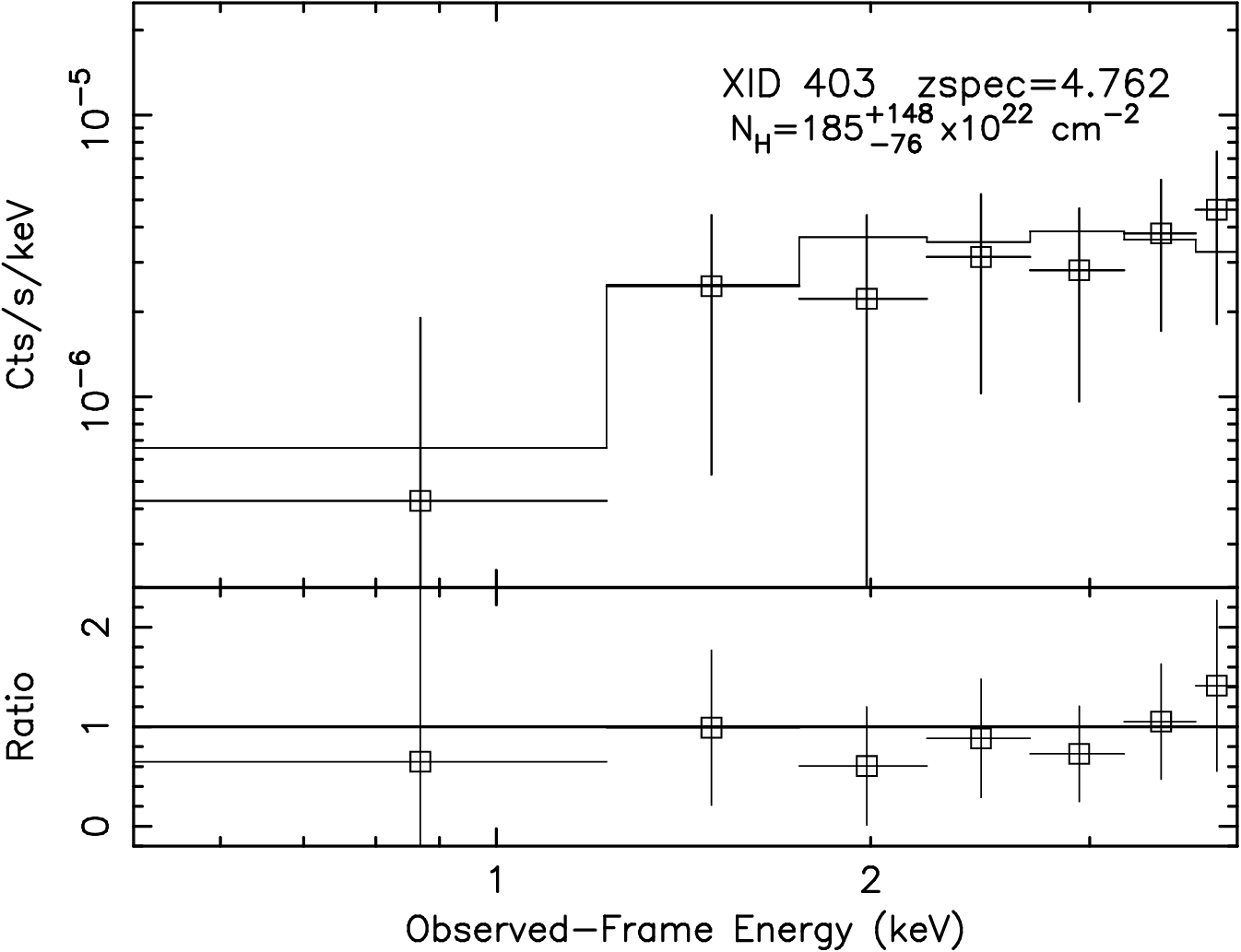}
 \includegraphics[width=57mm,height=75mm,angle=270]{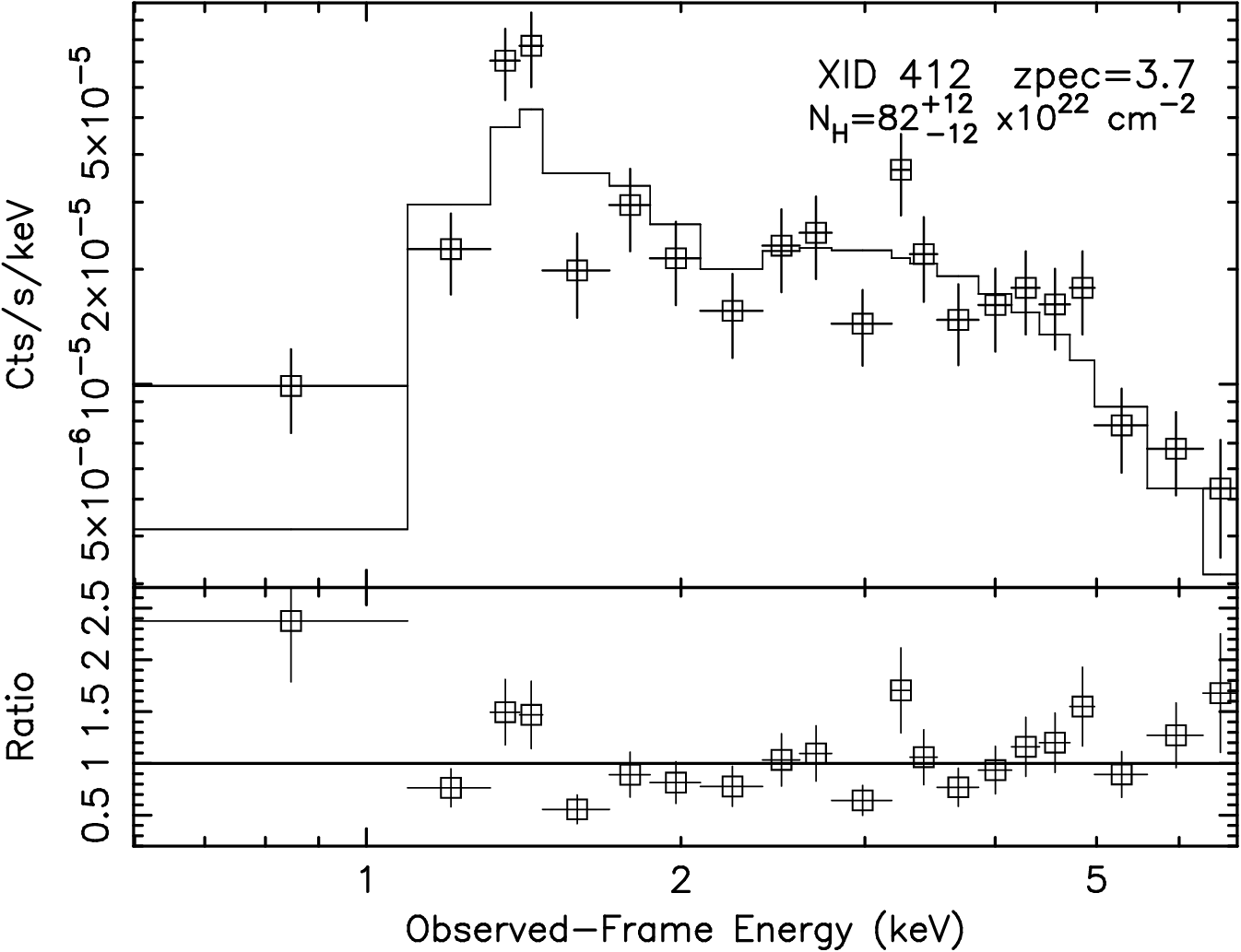}
 \includegraphics[width=57mm,height=75mm,angle=270]{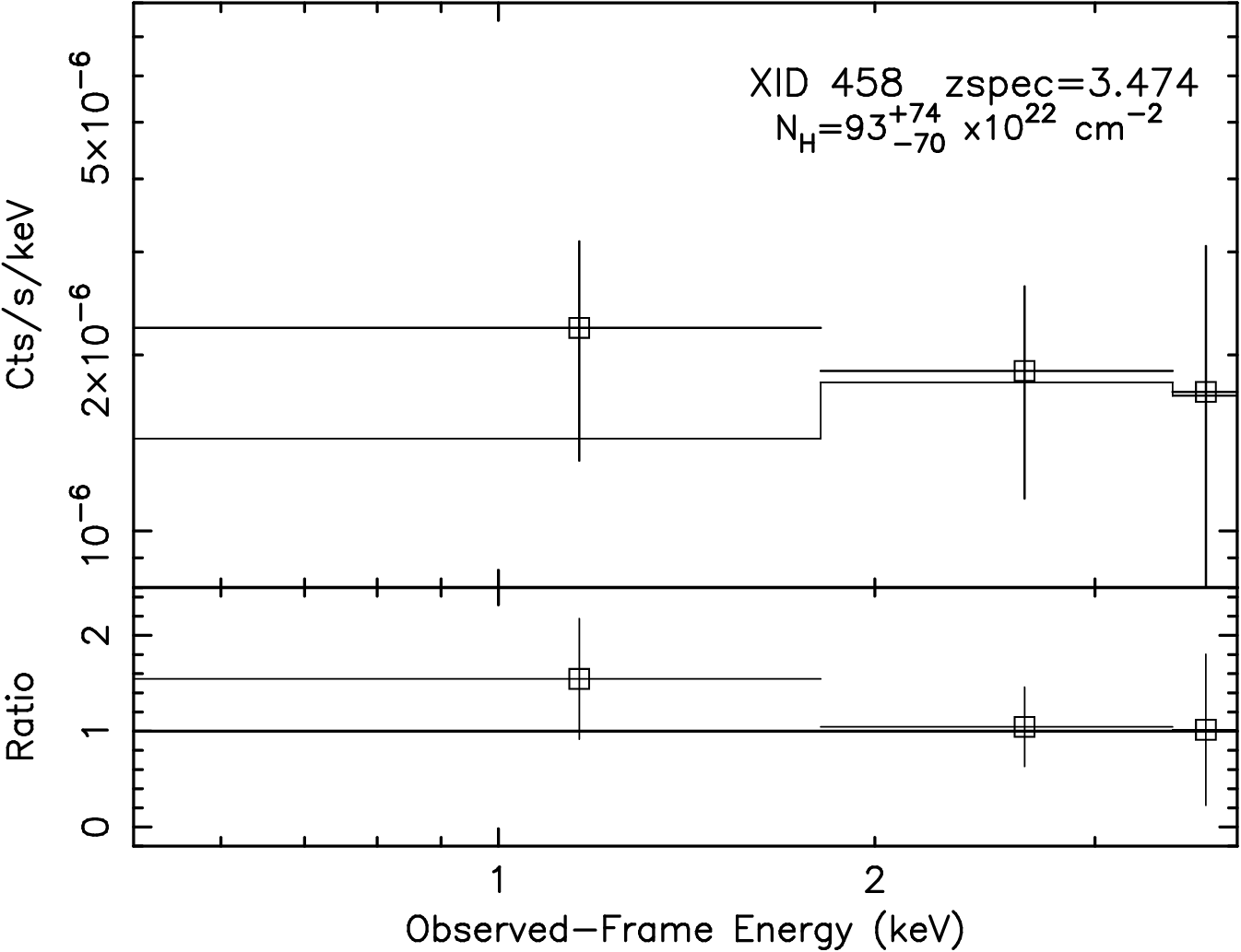}
\end{center}
\caption{(Continue)}
\end{figure*}

\addtocounter{figure}{-1}

\begin{figure*}
\begin{center}
 \includegraphics[width=57mm,height=75mm,angle=270]{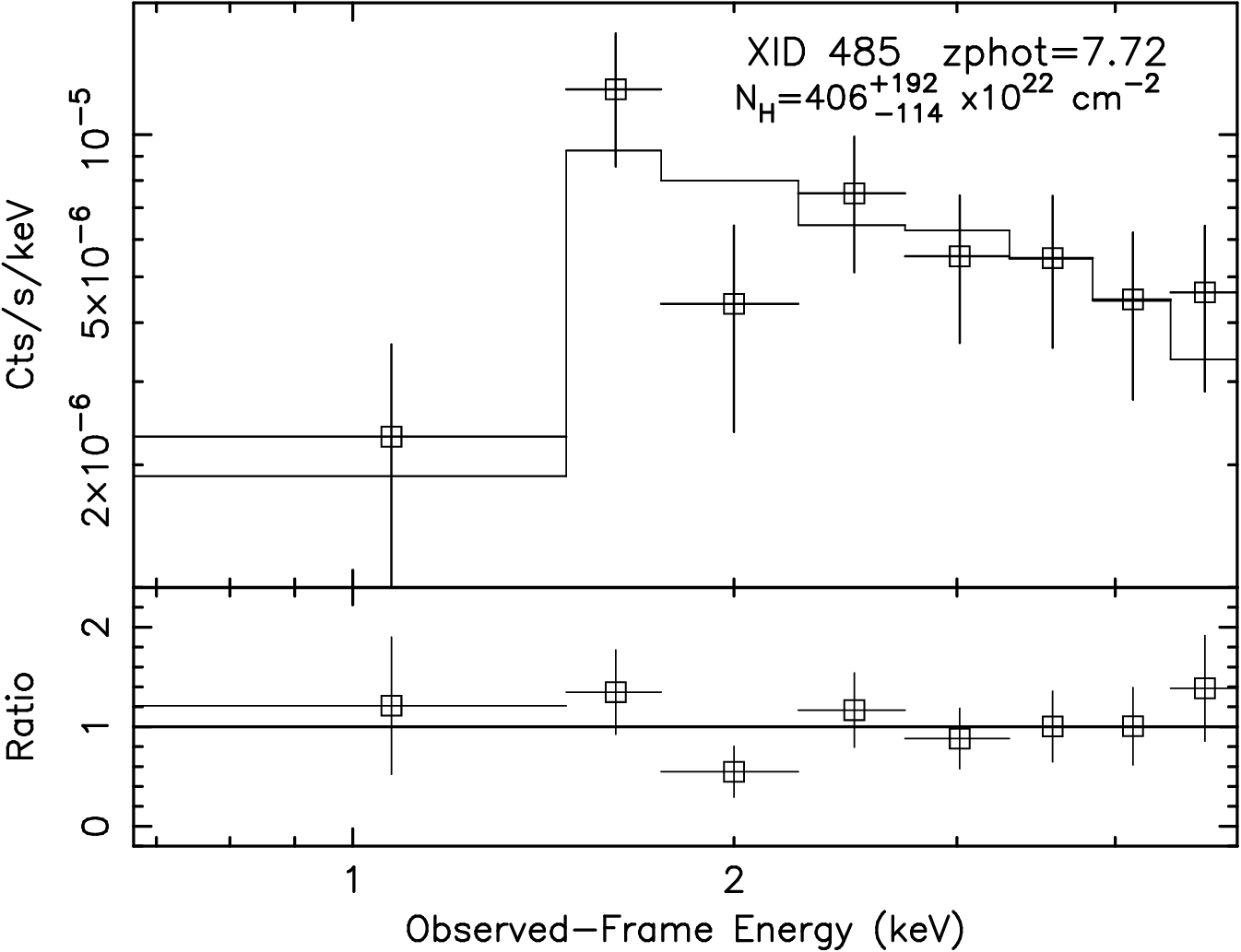}
 \includegraphics[width=57mm,height=75mm,angle=270]{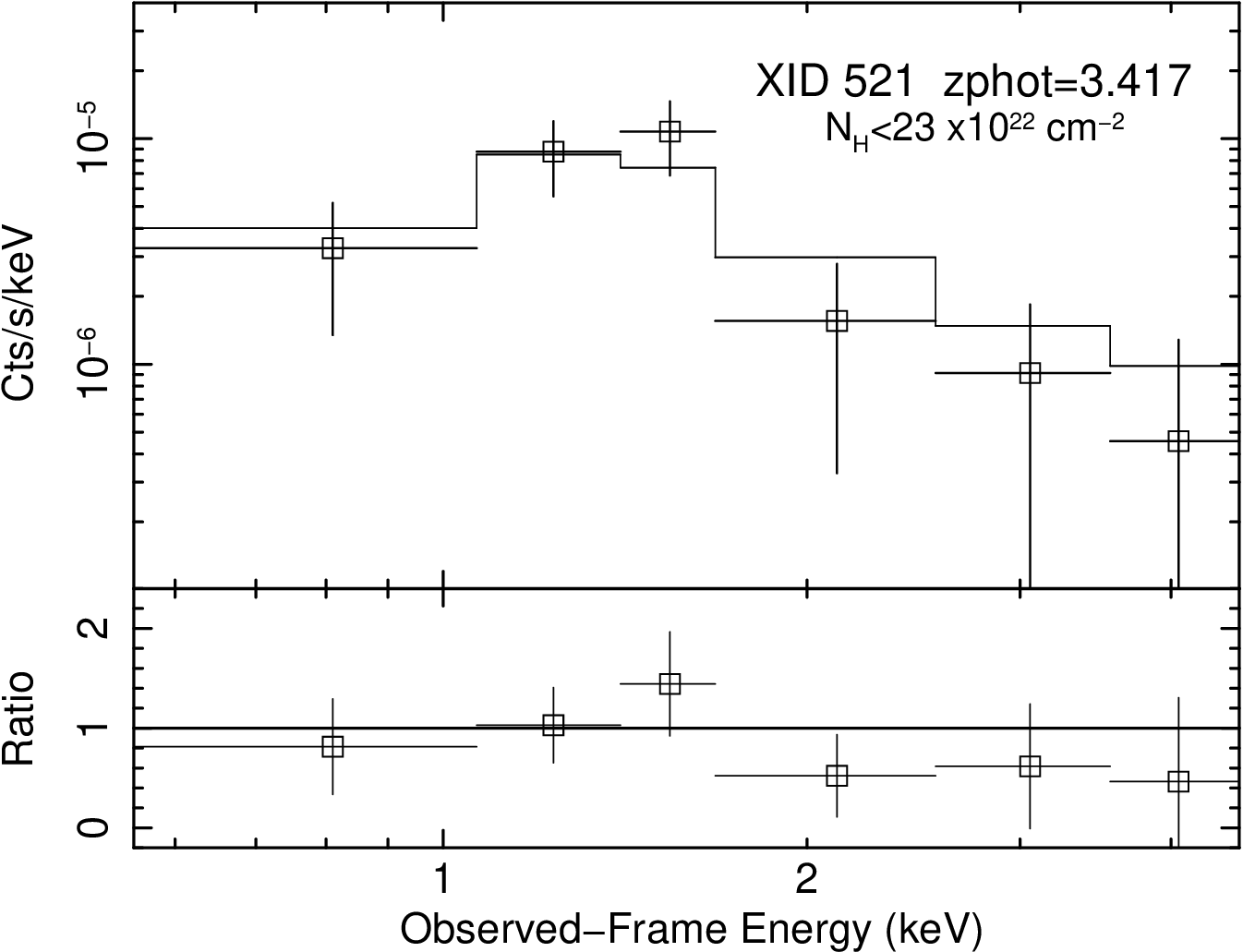}
 \includegraphics[width=57mm,height=75mm,angle=270]{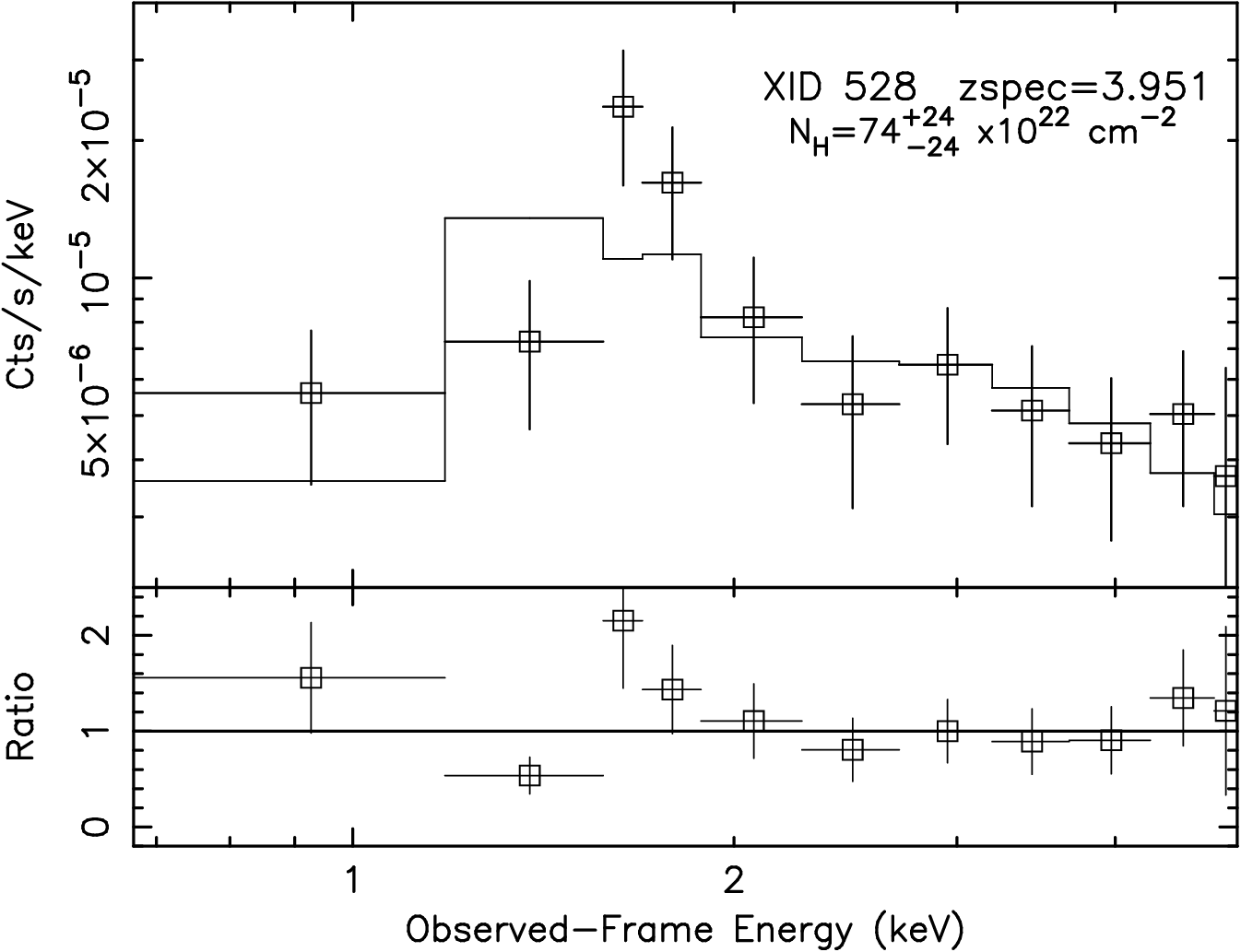}
  \includegraphics[width=57mm,height=75mm,angle=270]{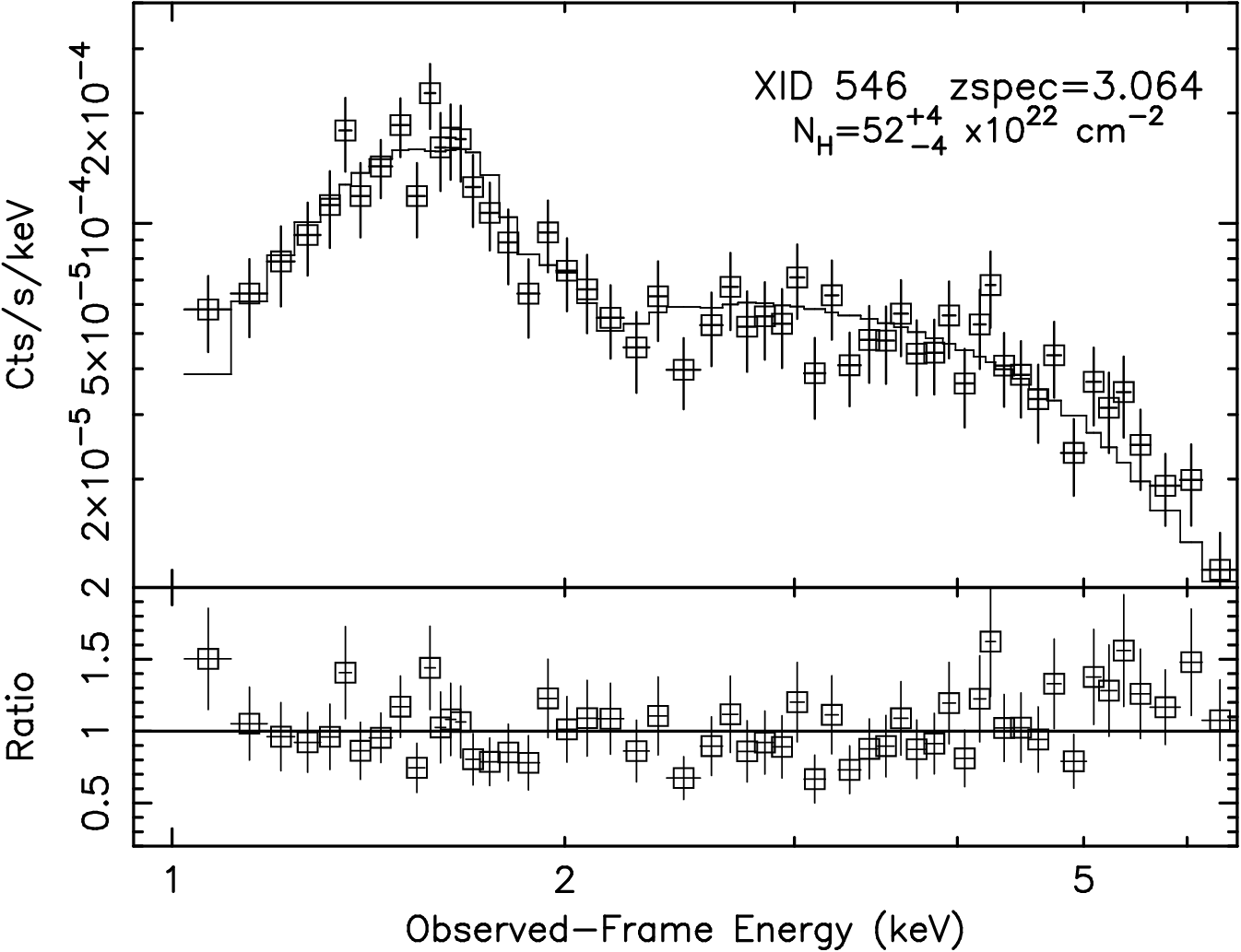}
 \includegraphics[width=57mm,height=75mm,angle=270]{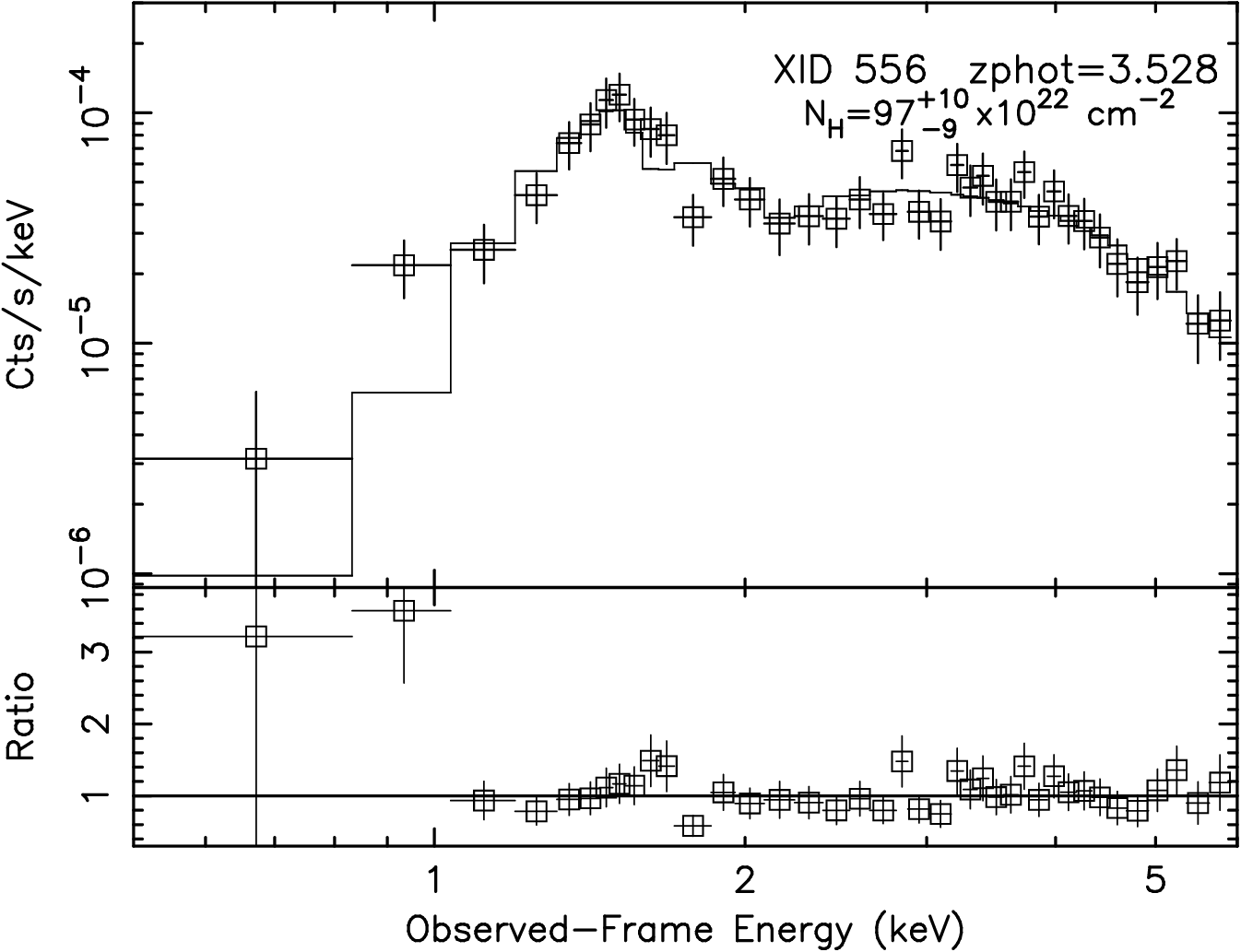}
 \includegraphics[width=57mm,height=75mm,angle=270]{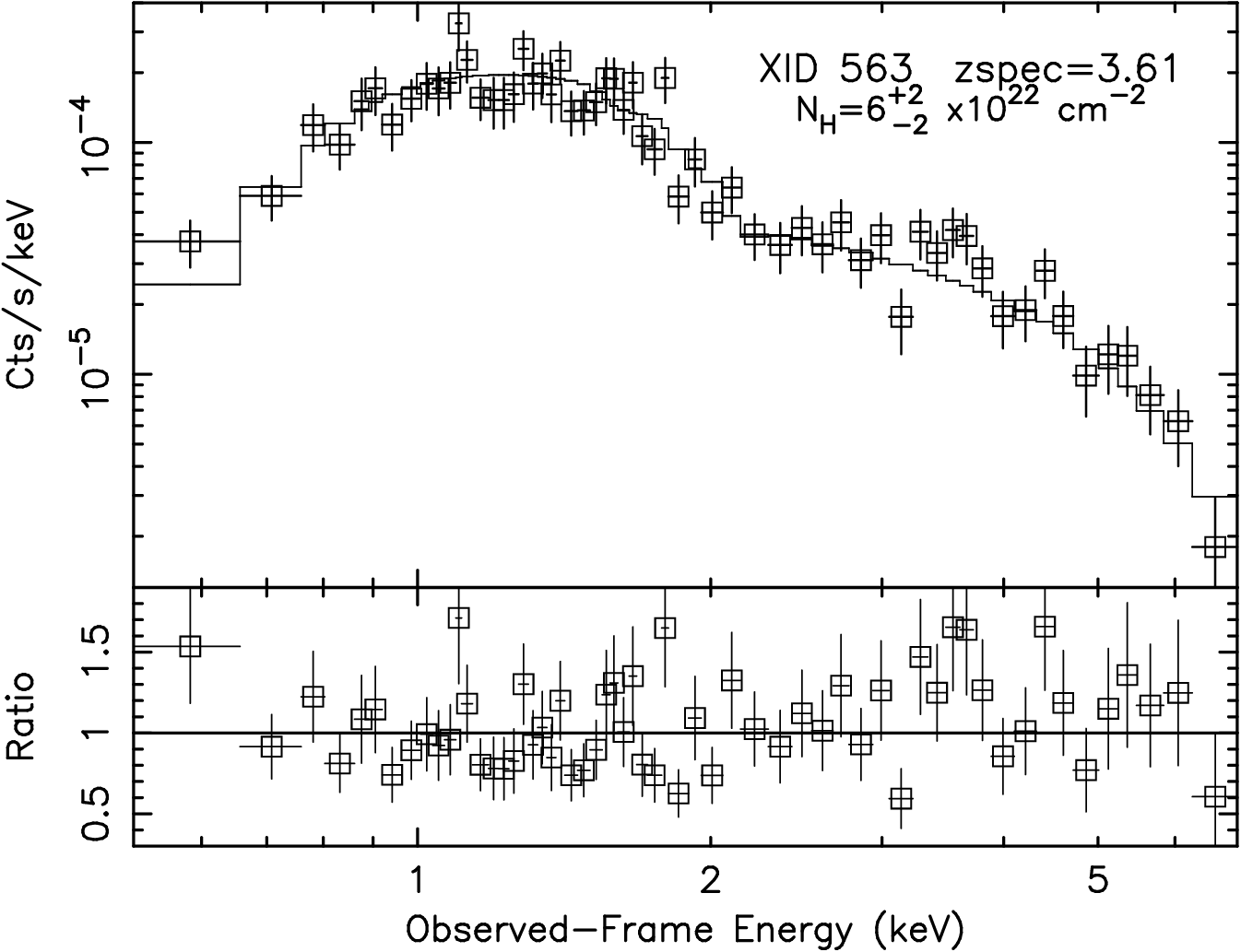}
 \includegraphics[width=57mm,height=75mm,angle=270]{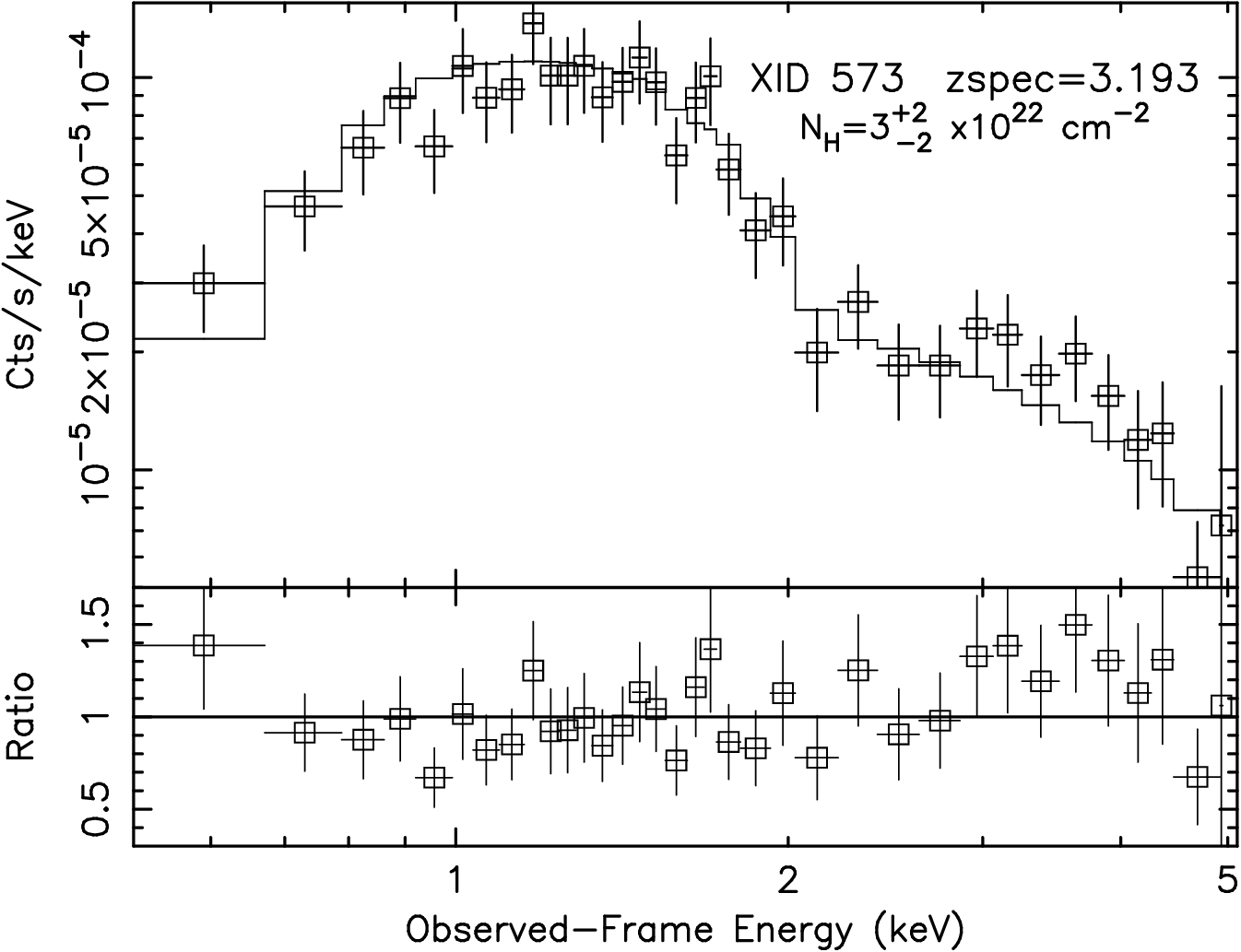}
 \includegraphics[width=57mm,height=75mm,angle=270]{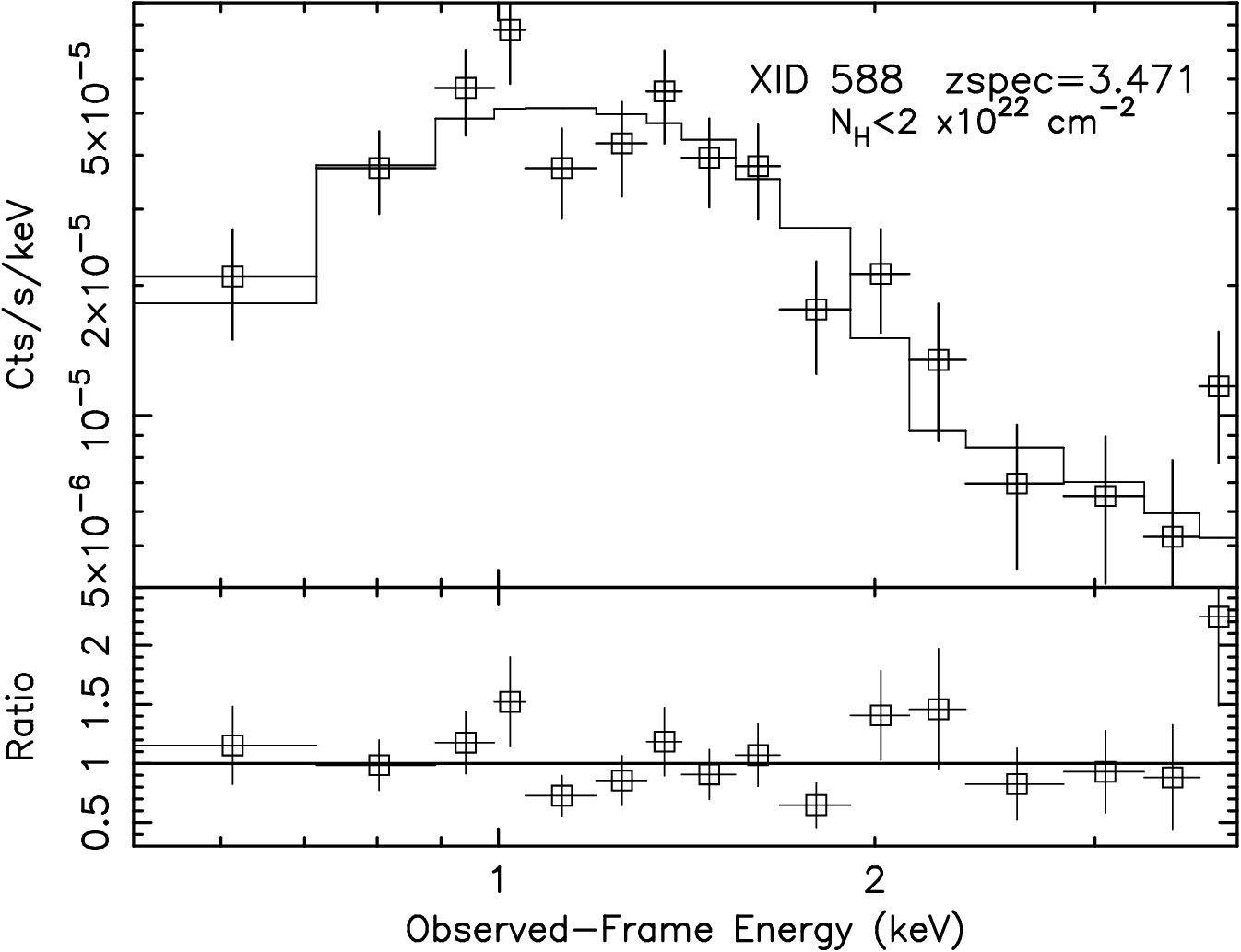}
\end{center}
\caption{(Continue)}
\end{figure*}

\addtocounter{figure}{-1}

\begin{figure*}
\begin{center}
 \includegraphics[width=57mm,height=75mm,angle=270]{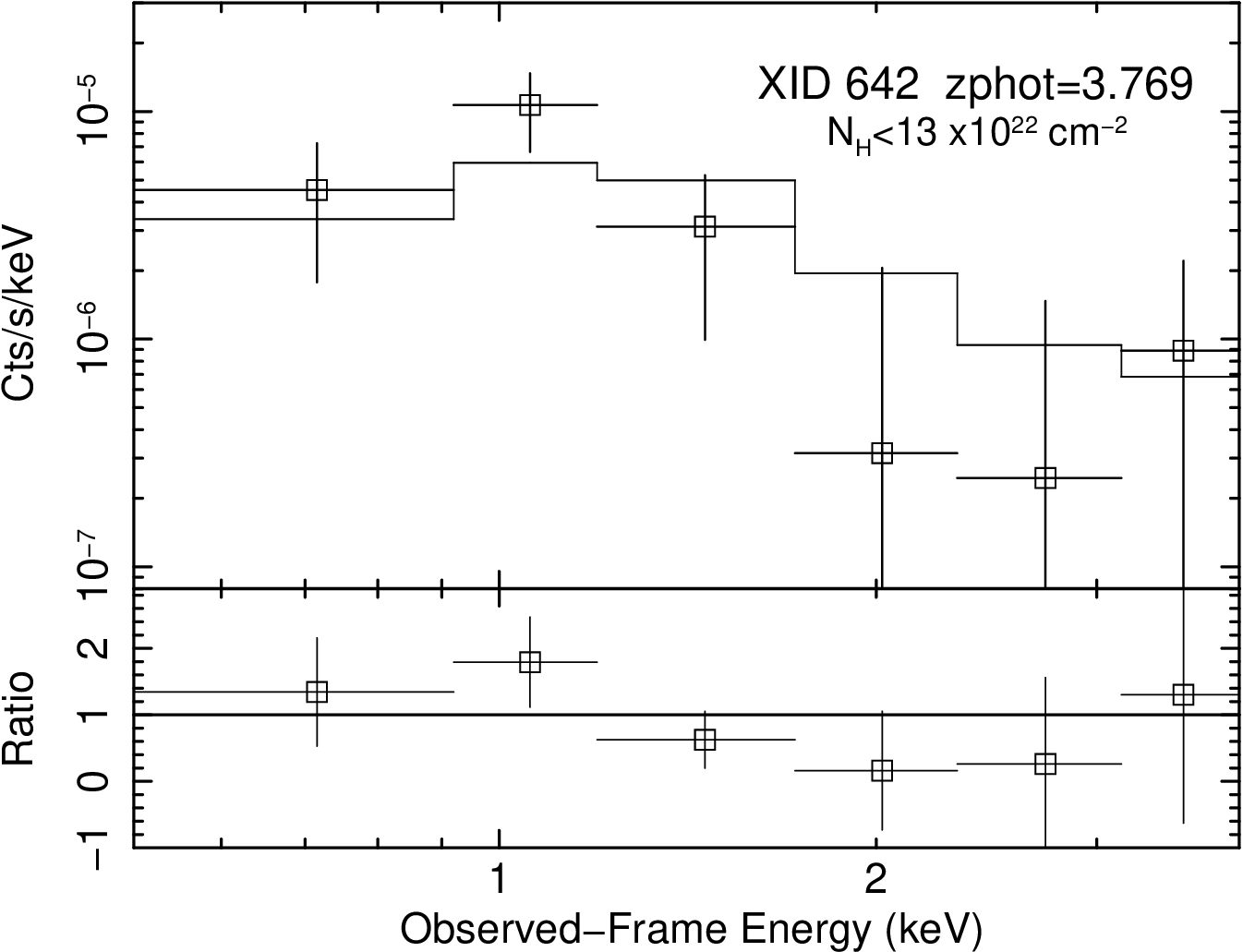}
 \includegraphics[width=57mm,height=75mm,angle=270]{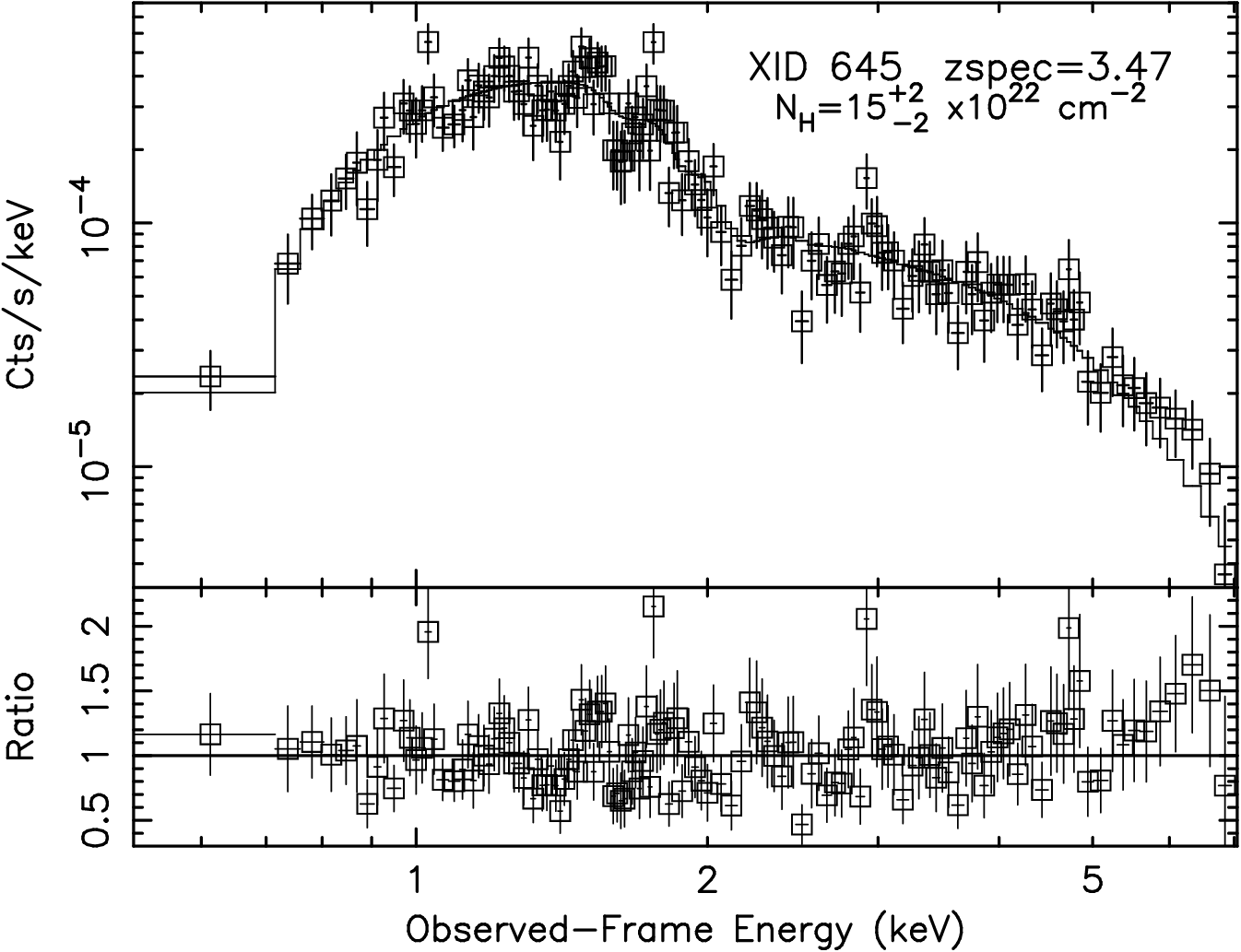}
 \includegraphics[width=57mm,height=75mm,angle=270]{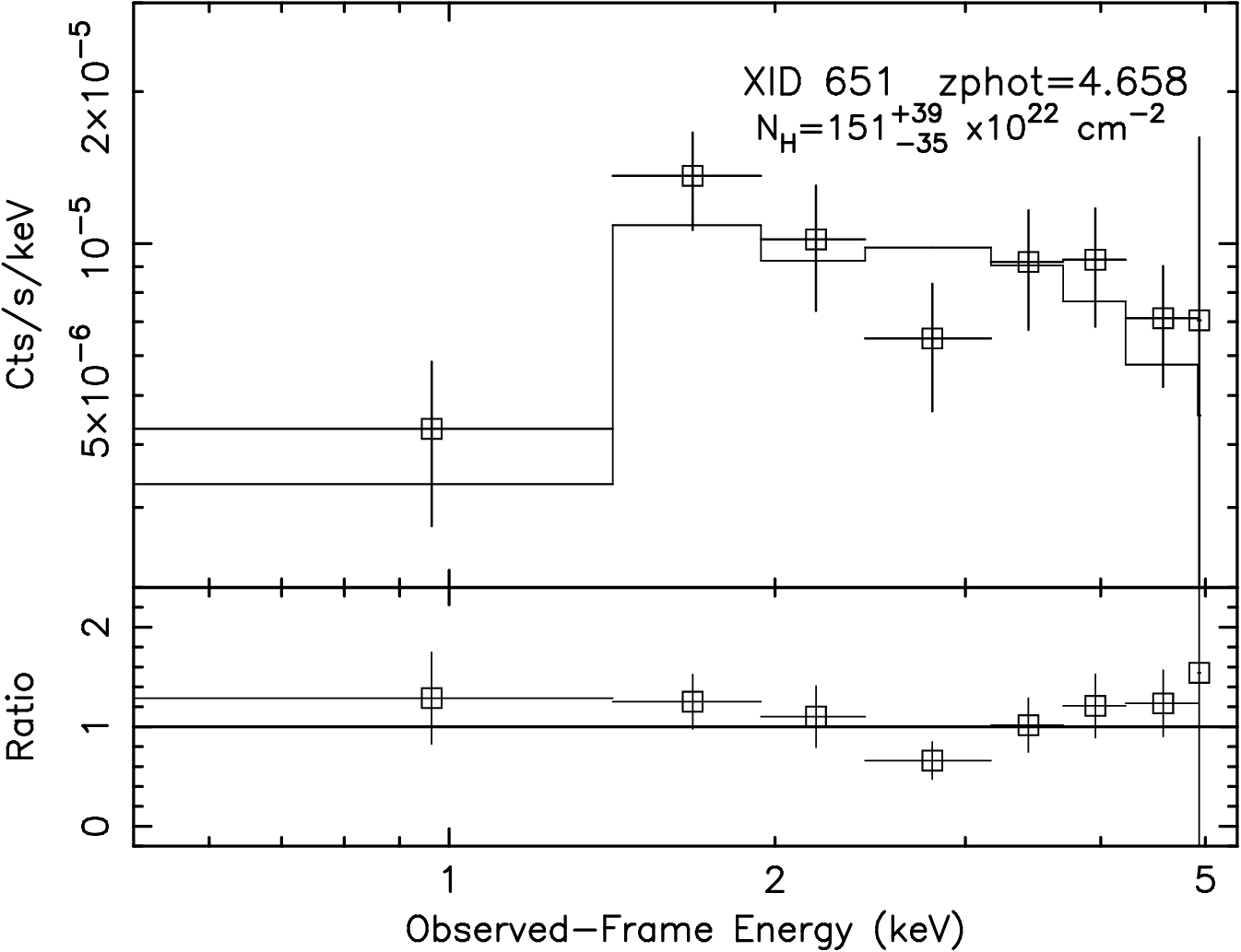}
  \includegraphics[width=57mm,height=75mm,angle=270]{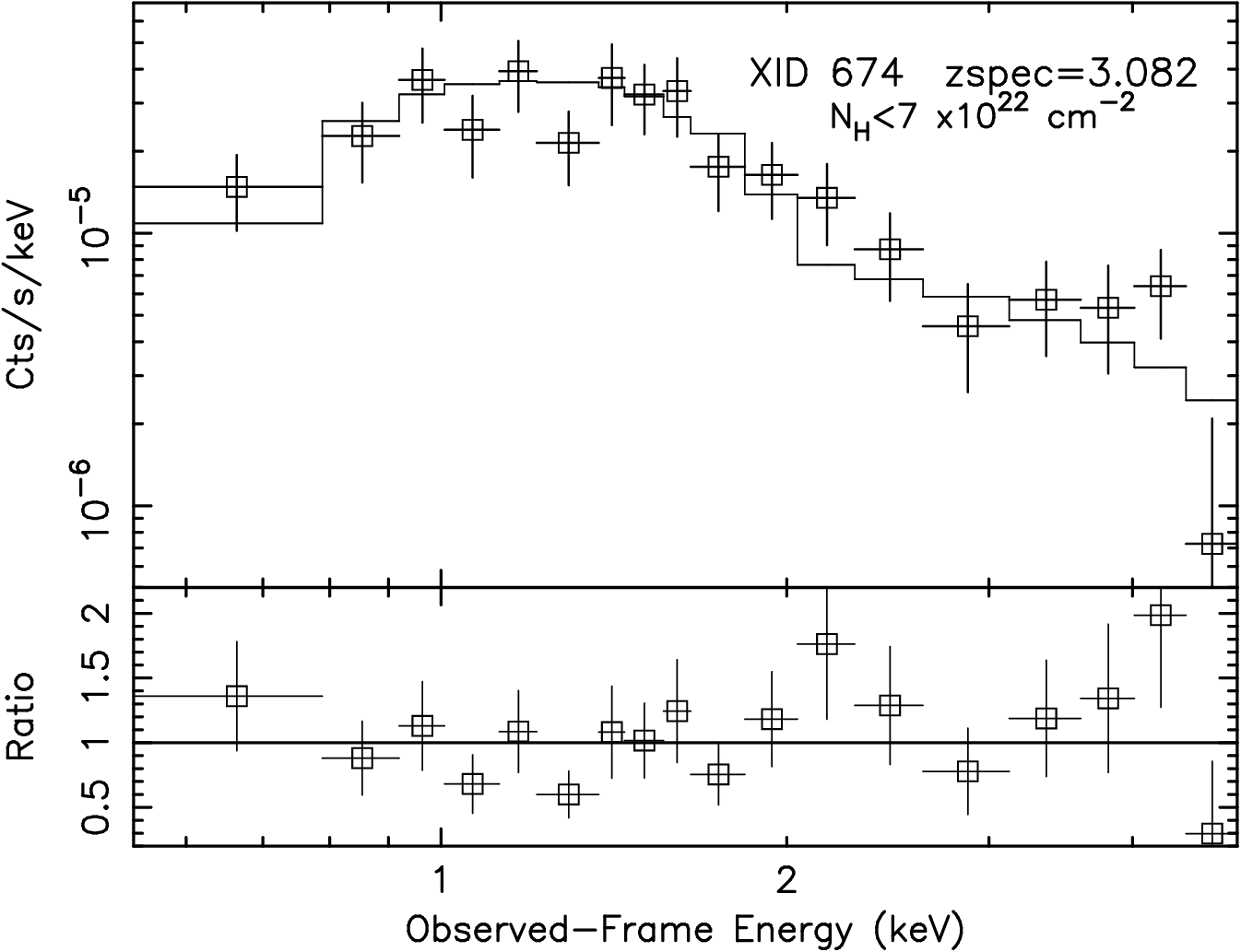}
 \includegraphics[width=57mm,height=75mm,angle=270]{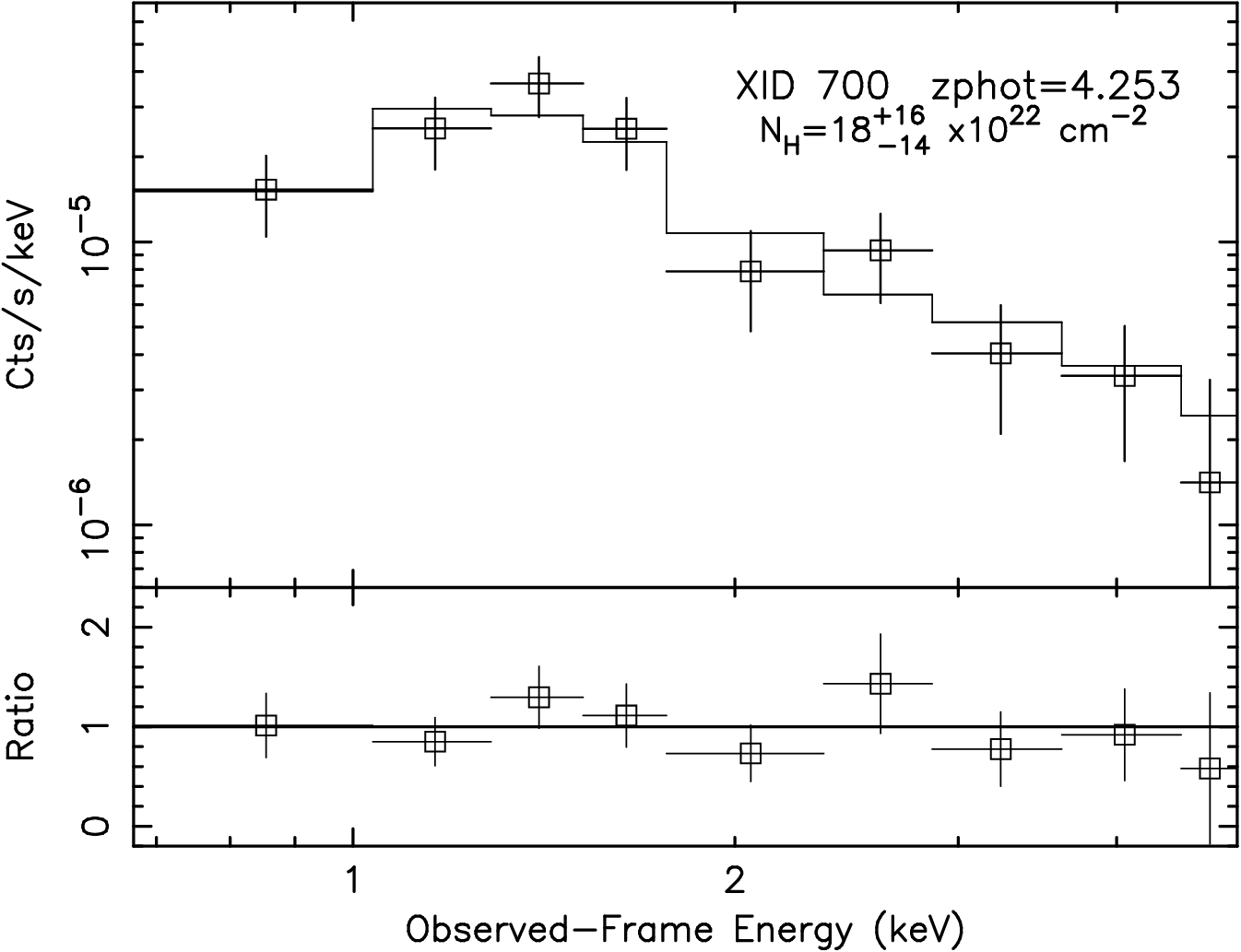}
 \includegraphics[width=57mm,height=75mm,angle=270]{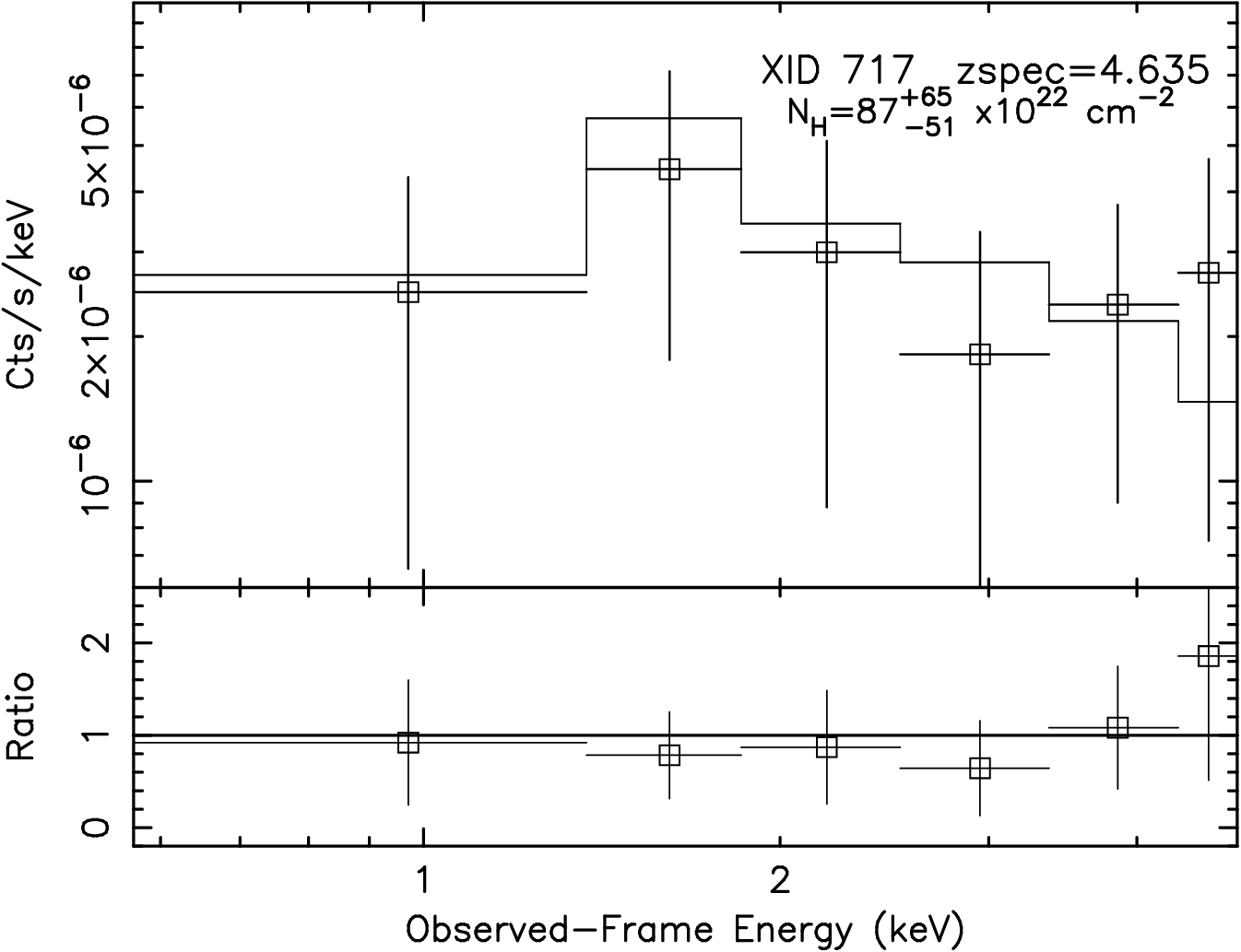}
 \includegraphics[width=57mm,height=75mm,angle=270]{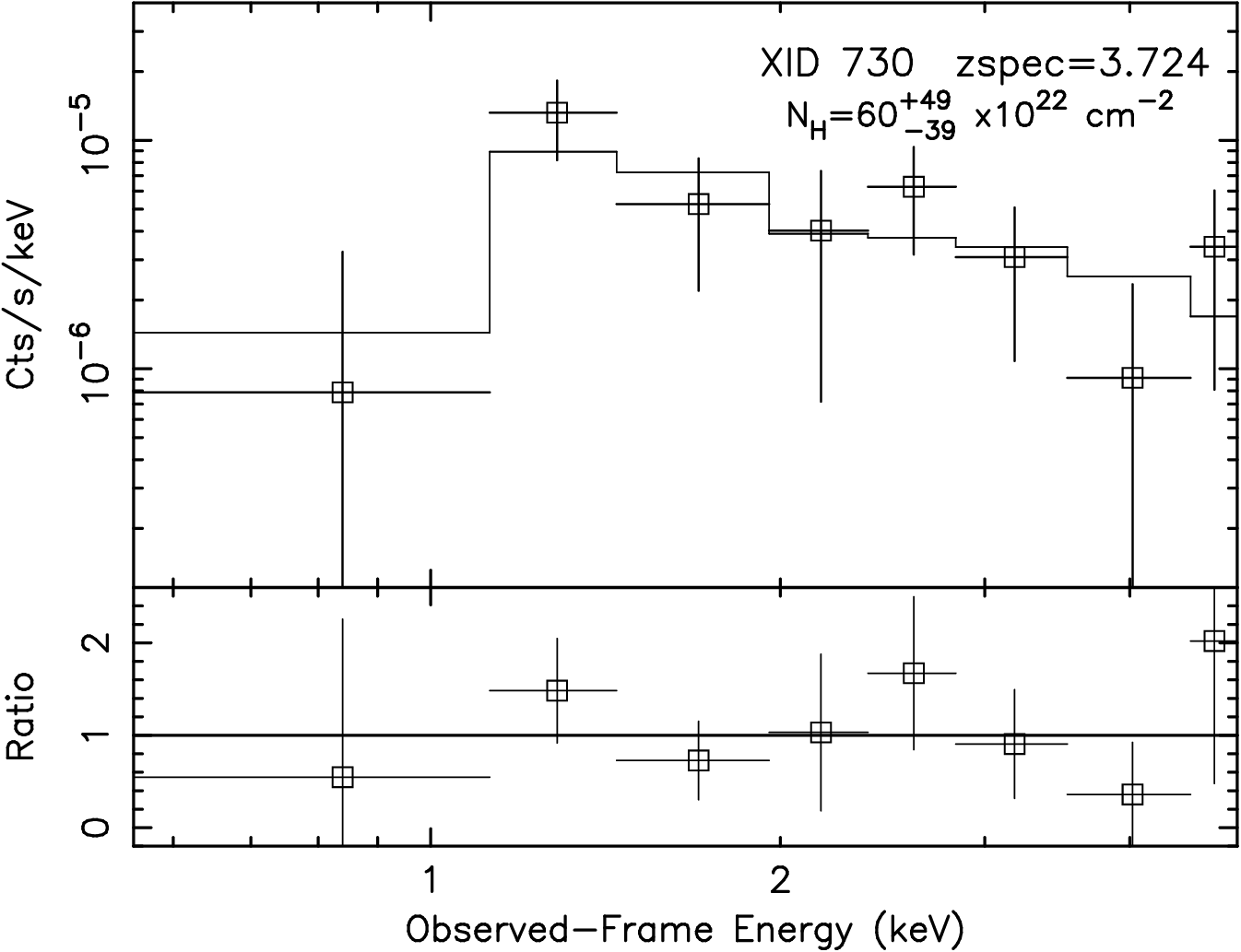}
\end{center}
\caption{(Continue)}
\end{figure*}

\label{lastpage}

\end{document}